\documentclass[a4paper,10pt]{article}
\usepackage[english]{babel}
\usepackage{jheppub}
\usepackage[T1]{fontenc} 
\usepackage{graphicx}
\usepackage{amsmath}
\usepackage{color}
\usepackage{xpatch}
\usepackage{caption}
\usepackage{subcaption}
\usepackage{slashed}
\usepackage{cleveref}
\usepackage{tikz}
\usepackage{multicol}
\usepackage{multirow}
\usepackage{hyperref}
\usepackage{slashed}
\usepackage{makecell}
\usetikzlibrary{decorations.pathreplacing}
\usepackage[normalem]{ulem}

\renewcommand{\i}{\ensuremath{\mathrm{i}}}
\renewcommand{\d}{\ensuremath{\mathrm{d}}}
\newcommand{\eps}{\ensuremath{\epsilon}}

\xpretocmd{\eqref}{eq.~}{}{}
\newcommand{\qb}{{\bar{q}}}
\newcommand{\tb}{{\bar{t}}}
\newcommand{\as}{{\alpha_s}}

\def\vareps{\varepsilon}
\newcommand{\GammaBold}{\mathbf{\Gamma}}
\newcommand{\GammaPrime}{\Gamma^\prime}
\def\qb{\bar{q}}
\def\cH{\mathcal{H}}
\def\cS{\mathcal{S}}
\def\cG{\mathcal{G}}
\def\cT{\mathcal{T}}
\def\cM{\mathcal{M}}
\def\cA{\mathcal{A}}
\def\cF{\mathcal{F}}
\def\cO{\mathcal{O}}
\def\cR{\mathcal{R}}
\def\ZZ{\mathbf{Z}}
\def\dZ{\delta Z}
\def\dk#1{\frac{\mathrm{d}^d k_{#1}}{\mathrm{i} \pi^{d/2} }}
\def\xx{x}
\def\pf{f}
\def\momperm{\vartheta}
\def\la{\langle}
\def\ra{\rangle}
\def\spA#1#2{\la#1#2\ra}
\def\spB#1#2{[#1#2]}
\def\spAB#1#2#3{\la#1|#2|#3]}

\def\spAA#1#2#3#4{\la#1|#2|#3|#4\ra}

\def\ttggg{0\to\bar{t}tggg}
\def\ttqqg{0\to\bar{t}t\bar{q}qg}

\def\colhelsum{\underset{\mathrm{colour}}{\overline{\sum}}\,\underset{\mathrm{pol.}}{\overline{\sum}}}

\def\mren{\mathrm{mren}}
\newcommand{\mtsq}{\ensuremath{m_{t}^2}}

\definecolor{mygreen}{rgb}{0,0.7,0}

\allowdisplaybreaks[4]

\title{Double virtual QCD corrections to $t\bar{t}+$jet production at the LHC}

\author[a]{Simon Badger,}
\author[b]{Matteo Becchetti,}
\author[a]{Colomba Brancaccio,}
\author[c]{Micha\l{} Czakon,}
\author[d,e]{Heribertus Bayu Hartanto,}
\author[f]{Rene Poncelet,}
\author[g]{Simone Zoia}
\affiliation[a]{Dipartimento di Fisica and Arnold-Regge Center, Università di Torino, and INFN, Sezione di Torino,
Via P.\ Giuria 1, I-10125 Torino, Italy}
\affiliation[b]{Dipartimento di Fisica e Astronomia, Università di Bologna e INFN, Sezione di Bologna, via Irnerio 46,
I-40126 Bologna, Italy}
\affiliation[c]{Institut f\"ur Theoretische Teilchenphysik und Kosmologie, RWTH Aachen University, D-52056 Aachen, Germany}
\affiliation[d]{Asia Pacific Center for Theoretical Physics, Pohang, 37673, Korea}
\affiliation[e]{Department of Physics, Pohang University of Science and Technology, Pohang, 37673, Korea}
\affiliation[f]{The Henryk Niewodnicza\'nski Institute of Nuclear Physics, ul.\ Radzikowskiego 152, 31-342 Krakow, Poland}
\affiliation[g]{Physik-Institut, Universität Zürich, Winterthurerstrasse 190, 8057 Zürich, Switzerland}

\emailAdd{simondavid.badger@unito.it, matteo.becchetti@unibo.it, colomba.brancaccio@unito.it, 
mczakon@physik.rwth-aachen.de, bayu.hartanto@apctp.org, rene.poncelet@ifj.edu.pl, simone.zoia@physik.uzh.ch}
 
\preprint{P3H-25-092, TTK-25-38, ZU-TH~74/25, IFJPAN-IV-2025-23}
 
\abstract{
We present a leading colour computation of the double virtual contributions to
top-quark pair production in association with a jet at a hadron collider at next-to-next-to-leading order in
QCD. The finite remainders of the two-loop amplitudes, after subtraction of
infrared and ultraviolet divergences, are extracted analytically from
evaluations over finite fields by using a (potentially) overcomplete basis of special
functions defined through their differential equations. We construct the colour-
and spin-summed interference with the tree-level amplitudes and present a
\texttt{C++} library suitable for immediate use in phenomenological studies. We present
new techniques for the evaluation of the special functions through direct
numerical integration of differential equations which perform well across
the full physical phase space.
}

\date{}
\begin{document}
\maketitle
\flushbottom

\section{Introduction}

The top quark physics program at the LHC continues to produce a wealth of
experimental data that can be used to study the fundamental parameters of the
Standard Model (SM). Precise theoretical predictions for the dominant production
channels are in high demand as experimental uncertainties continue to drop,
requiring computations at least at next-to-next-leading order (NNLO) in the strong
coupling.

Studies of top quark pair production are by now well underway with NNLO QCD and
NLO electroweak (EW) corrections already being used to compare with experimental data, see for example refs.~\cite{CMS:2018htd,ATLAS:2023gsl}.
Precise predictions for $t\bar{t}+$jet
production are also in high demand owing to the high number (around 50\%) of
$t\bar{t}$ events that come with an associated jet and have been studied by both CMS and ATLAS experiments~\cite{CMS:2016oae,ATLAS:2018acq,CMS:2020grm,CMS:2024ybg}. This process has been
available at NLO in QCD (with on-shell top quarks) since
2007~\cite{Dittmaier:2007wz} and is now available via a number of automated
tools which offer a full set of decays, EW corrections, parton shower and
off-shell effects~\cite{Alioli:2011as,Melnikov:2011qx,Bevilacqua:2015qha,Bevilacqua:2016jfk,G_tschow_2018}. It has also been noted that the normalised
distributions for $t\bar{t}j$ are highly sensitive to the top-quark mass~\cite{Alioli:2013mxa,Bevilacqua:2017ipv,Alioli:2022ttk,Alioli:2022lqo}, offering an alternative method for determining the top-quark mass, which has already been employed by both the CMS~\cite{CMS:2022emx} and ATLAS~\cite{ATLAS:2025ona} collaboration. Over the last 10 years or so there has been substantial progress in
computational methods for two-to-three scattering amplitudes, where new
techniques have been required to overcome the challenges of both analytic
(Feynman integrals and special functions) and algebraic (kinematic
coefficients) complexity. In particular development of highly optimised
integration-by-parts reduction
techniques~\cite{Gluza:2010ws,Ita:2015tya,Larsen:2015ped,Wu:2023upw,Guan:2024byi} and
improved analytic understanding of multi-scale Feynman integrals through their
differential
equations~\cite{Gehrmann:2015bfy,Papadopoulos:2015jft,Abreu:2018rcw,Chicherin:2018mue,Chicherin:2018old,Abreu:2018aqd,Abreu:2020jxa,Canko:2020ylt,Abreu:2021smk,Kardos:2022tpo,Abreu:2023rco} (DEs)
have had a dramatic impact. It has been possible to classify the integrals for
processes with massless internal propagators and five massless external legs or
four massless and one massive external leg in terms of a basis so-called
`pentagon
functions'~\cite{Gehrmann:2018yef,Chicherin:2020oor,Chicherin:2021dyp,Abreu:2023rco} (see also refs.~\cite{Badger:2023xtl,FebresCordero:2023pww,Gehrmann:2024tds}).
These functions provide a basis for the infrared (IR) and ultraviolet (UV)
subtracted finite remainders which can be evaluated efficiently during
phase-space integration.  The rational coefficients multiplying these special
functions can be extracted from numerical tensor integral reduction over finite
fields, sidestepping the large intermediate expressions that appear in
traditional
approaches~\cite{vonManteuffel:2014ixa,Peraro:2016wsq,Klappert:2019emp,Peraro:2019svx,Smirnov:2019qkx,Klappert:2020aqs,Klappert:2020nbg}.
This program has resulted in an impressive number of new results for double
virtual QCD corrections to processes with massless internal
particles~\cite{Badger:2019djh,Badger:2021nhg,Badger:2021ega,Abreu:2021asb,Badger:2022ncb,Agarwal:2021vdh,Badger:2021imn,Abreu:2023bdp,Badger:2023mgf,Agarwal:2023suw,DeLaurentis:2023nss,DeLaurentis:2023izi},
with up to one additional external
scale~\cite{Badger:2021nhg,Badger:2021ega,Abreu:2021asb,Badger:2022ncb,Badger:2024sqv,Badger:2024mir,DeLaurentis:2025dxw}
and including internal massive propagators~\cite{Agarwal:2024jyq}. Many of
these results have already been combined with contributions for unresolved
radiation to make new precise predictions for differential cross
sections~\cite{Chawdhry:2019bji,Kallweit:2020gcp,Chawdhry:2021hkp,Czakon:2021mjy,Badger:2021ohm,Chen:2022ktf,Alvarez:2023fhi,Badger:2023mgf,Hartanto:2022qhh,Hartanto:2022ypo,Buonocore:2022pqq,Catani:2022mfv,Buonocore:2023ljm,Mazzitelli:2024ura,Devoto:2024nhl,Biello:2024pgo,Buccioni:2025bkl}.

A key ingredient in the development of the pentagon functions approach to solving Feynman integrals was the possibility of constructing a canonical system of DEs~\cite{Henn:2013pwa} described by logarithmic one-forms. In this framework, Feynman integrals are typically represented by polylogarithmic functions, whose analytic properties and numerical evaluation are well understood. A major challenge in computing processes with massive internal propagators is the emergence of special functions associated with elliptic curves and higher-genus geometries~\cite{Bourjaily:2022bwx}. In this case a solution along
the lines of the pentagon functions is currently not available. In particular, while canonical DEs can also be constructed in this context, the numerical evaluation of their solutions remains an open problem.  In previous studies, it was shown that
even at leading colour the two-loop integral basis for $t\bar{t}+$jet
production contained elliptic structures that block the established path to the
evaluation of the special functions~\cite{Badger:2022hno,Badger:2024fgb,Becchetti:2025oyb}.
Nevertheless, it was possible to circumvent the difficulties associated with elliptic structures by identifying a basis of integrals in which the contributions from elliptic integrals sectors appear only in the finite remainder. While this basis of integrals does not satisfy canonical DEs, this approach opened up the possibility of
constructing a set of special functions in which redundancies are reduced
enough to enable the analytic extraction of the finite remainder while also
yielding important simplifications in its expression, although some functional
relations may be missed.  This (potentially) overcomplete basis of special
functions is defined and evaluated as the solution to a system of DEs which contains only algebraic functions.  As a
proof of concept, it was shown in ref.~\cite{Badger:2024dxo} that this method
was good enough to enable the first numerical evaluation of the leading colour
two-loop amplitudes in the gluon channel.

In this article we complete the analytic reconstruction of the leading-colour two-loop helicity amplitudes in all relevant channels using the previously identified special function basis and finite-field reconstruction techniques. These amplitudes retain the full spin information, enabling the treatment of top-quark decays. The amplitudes are subsequently inserted into the colour- and spin-summed double virtual hard function, for which we must include additional permutations of the Feynman integral topologies. We construct a system of differential equations for the full set of special functions contributing to the hard function, and present a new strategy for its numerical integration.

The evaluation of Feynman integrals (or the special functions that they may be
written in terms of) through numerical integration of their differential
equations has been a successful method in the past, with its first applications in refs.~\cite{Boughezal:2007ny,Czakon:2008zk}. In this article we
present a modified strategy able to handle new features appearing in the
differential equation, specifically the appearance of square roots as well as the large size of the system, for the
special function basis of the hard function. This strategy to evaluate the hard
functions is implemented in a \texttt{C++} library such that our results
are ready for immediate deployment in phenomenological studies~\cite{Badger:2025ilt}.

Our article is organised as follows: We first detail the colour decomposition
into partial amplitudes, classification of ultraviolet and infrared
singularities and definition of the hard function in \cref{sec:partial}.
We then describe the computation of the helicity amplitudes using the four-dimensional 
projector method and integral reduction over finite fields in
\cref{sec:helamps}. Special functions and the
new evaluation strategy are discussed in \cref{sec:specfuncs}. We then
present some benchmark results and cross checks together with the \texttt{C++} implementation of the amplitudes in \cref{sec:res} before
reaching our conclusions. Analytic results and the \texttt{C++} implementation of the hard functions are
included with the ancillary files~\cite{zenodo}. 

\section{Partial amplitude decomposition and pole structure \label{sec:partial}}

In this section we define our notation and describe the colour structure as well as the ultraviolet (UV) and infrared (IR) pole structure of the relevant scattering amplitudes. 
We compute two-loop QCD corrections to the scattering amplitudes of the following partonic processes:
\begin{subequations}
\label{eq:processdefinition}
\begin{align} 
	 & 0 \to \bar{t}(p_1) + t(p_2) + g(p_3) + g(p_4) + g(p_5)  \,, \\
	 & 0 \to \bar{t}(p_1) + t(p_2) + \bar{q}(p_3) + q(p_4) + g(p_5) \,.
\end{align}
\end{subequations}
All other partonic channels relevant to $t\bar{t}$+jet production at hadron colliders can be derived from the two above by permuting the external momenta.

The external momenta are taken to be outgoing, and satisfy momentum conservation
\begin{equation}
	\sum_{i=1}^{5} p_i = 0 \,, 
\end{equation}
and on-shell conditions
\begin{align} \label{eq:onshell}
	&  p_1^2 = \mtsq \,, \qquad 
	p_2^2 = \mtsq \,, \qquad
	p_3^2 = 0 \,, \qquad
	p_4^2 = 0 \,, \qquad
	p_5^2 = 0 \,,
\end{align}
where \(m_t\) is the top-quark mass.
The kinematics of the process are described by the following six scalar invariants,
\begin{align}
\label{eq:dijmtset}
\vec{d} = \bigl( d_{12}, d_{23}, d_{34}, d_{45}, d_{15}, m_t^2 \bigr) \,,
\end{align}
with $d_{ij} = p_i \cdot p_j$, together with a pseudo-scalar invariant,
\begin{align}
\label{eq:tr5definition}
\begin{aligned}
	\mathrm{tr}_5 = \mathrm{tr}\big(\gamma_5 \slashed{p}_1 \slashed{p}_2 \slashed{p}_3 \slashed{p}_4 \big)  = 4 \, \i \, \varepsilon_{\mu\nu\rho\sigma} \, p_1^{\mu} p_2^{\nu} p_3^{\rho} p_4^{\sigma} \,,
\end{aligned}
\end{align}
where $\varepsilon_{\mu\nu\rho\sigma}$ is the anti-symmetric Levi-Civita pseudo-tensor.
We work in the `t Hooft-Veltman (tHV) scheme: the external momenta are kept in four dimensions, while the loop momenta are treated in $d=4-2\eps$ dimensions.

We adopt the leading colour approximation, i.e.\ we consider only the leading contributions in the number of colour ($N_c$) and the number of closed massless fermion loops ($n_f$). 
We do not take into account contributions coming from loops of heavy quarks. 
In this approximation, the $L$-loop bare amplitudes have the following colour decomposition in terms of partial amplitudes:
\begin{align} \label{eq:colourdecompositionttggg}
&\begin{aligned}
	\cM^{(L)}\left( 1_{\bar{t}}, 2_t, 3_g,4_g,5_g \right) =& \; \sqrt{2} \, \bar{g}_s^3 \bigg[ (4\pi)^\eps e^{-\eps \gamma_E} \frac{\bar\alpha_s}{4\pi} \bigg]^L   \\
	& \times \sum_{\sigma \in Z_3} \left( t^{a_{\sigma(3)}} t^{a_{\sigma(4)}} t^{a_{\sigma(5)}} \right)_{i_2}^{\;\bar{i}_1} \,
	\cA^{(L)}_{g}\left( 1_{\bar{t}}, 2_t, \sigma(3)_g, \sigma(4)_g, \sigma(5)_g \right)  \,,
\end{aligned} \\
 \label{eq:colourdecompositionttqqg}
&\begin{aligned}
	\cM^{(L)}\left( 1_{\bar{t}}, 2_t, 3_{\qb},4_q,5_g \right) =& \; \sqrt{2} \, \bar{g}_s^3 \bigg[ (4\pi)^\eps e^{-\eps \gamma_E} \frac{\bar\alpha_s}{4\pi} \bigg]^L \times
          \bigg\lbrace \delta_{i_4}^{\;\bar{i}_1} \left( t^{a_5} \right)_{i_2}^{\;\bar{i}_3} \,
          \cA^{(L)}_{q}\left( 1_{\bar{t}}, 2_t, 3_{\qb}, 4_q, 5_g \right)  \\
        & \qquad + \delta_{i_2}^{\;\bar{i}_3} \left( t^{a_5} \right)_{i_4}^{\;\bar{i}_1} \,
	  \cA^{(L)}_{q}\left( 2_{\bar{t}}, 1_t, 4_{\qb}, 3_q, 5_g \right)  \bigg\rbrace \,,
\end{aligned}
\end{align}
for $\ttggg$ and $\ttqqg$, respectively.
Here, $t^{a}$ are the generators of the $su(N_c)$ colour algebra in the fundamental representation, normalised according to 
$\mathrm{tr} ( t^a t^b ) = \delta^{ab}/2$, $\bar{g}_s$ is the bare strong coupling constant with $\bar{\alpha}_s = \bar{g}_s^2/(4\pi)$, and $Z_3$ is the cyclic group of permutations of $\left\{3,4,5\right\}$.
All partial amplitudes in \cref{eq:colourdecompositionttggg,eq:colourdecompositionttqqg} can be obtained from $\cA^{(L)}_{g}\left( 1_{\bar{t}}, 2_t, 3_g, 4_g, 5_g \right)$ and $\cA^{(L)}_{q}\left( 1_{\bar{t}}, 2_t, 3_{\qb}, 4_q, 5_g \right)$ by suitable permutations. 
We set the renormalisation scale $\mu_R$ to $1$ throughout the computation of the partial amplitudes, and restore the dependence on it as discussed in \cref{app:mudep}

The partial amplitudes are in turn decomposed according to their $N_c$ and $n_f$ contributions, as 
\begin{subequations}
\label{eq:NcNfdecomposition}
\begin{align}
	\cA^{(1)}_{\xx} & = N_c \, A^{(1),N_c}_{\xx} + n_f \, A^{(1),n_f}_{\xx}  \,, \\
	\cA^{(2)}_{\xx} & = N_c^2 \, A^{(2),N_c^2}_{\xx} + N_c n_f \, A^{(2),N_c n_f}_{\xx} + n_f^2 \, A^{(2),n_f^2}_{\xx}  \,.
\end{align}
\end{subequations}
Hereinafter, we use $\xx \in \lbrace g, q \rbrace$ to label generically the $\ttggg$ and $\ttqqg$ partial amplitudes.
The decomposition in \cref{eq:NcNfdecomposition} holds also for the mass-renormalised and fully renormalised partial amplitudes defined below.

Bare scattering amplitudes for processes involving massive fermions are not gauge invariant. 
In order to work with gauge-invariant objects, we first define mass-renormalised amplitudes, which satisfy Ward identities,
by including the massive-quark mass counterterms, as
\begin{subequations}
\label{eq:massrenormalised}
\begin{align} 
        \cA^{(1)}_{\xx;\mathrm{mren}} & = \cA^{(1)}_\xx - \dZ_m^{(1)} \cA^{(0)}_{\xx;\mathrm{mct}} \,, \\
        \cA^{(2)}_{\xx;\mathrm{mren}} & = \cA^{(2)}_\xx - \dZ_m^{(2)} \cA^{(0)}_{\xx;\mathrm{mct}}
        + (\dZ_m^{(1)})^2 \cA^{(0)}_{\xx;2\mathrm{mct}} - \dZ_m^{(1)} \cA^{(1)}_{\xx;\mathrm{mct}} \,,
\end{align}
\end{subequations}
where $\cA^{(0)}_{\xx;\mathrm{mct}}$ ($\cA^{(0)}_{\xx;2\mathrm{mct}}$) is the tree-level partial amplitude with a single (double) mass-counterterm insertion, and $\cA^{(1)}_{\xx;\mathrm{mct}}$ is the one-loop partial amplitude with the insertion of a single mass counterterm (see also ref.~\cite{Badger:2024dxo} for further details).
We then include the UV counterterms originating from the renormalisation of the top-quark wave function and the strong coupling constant,
which leads to the fully renormalised amplitudes that are both gauge invariant and free of UV singularities, as 
\begin{subequations}
\label{eq:UVrenormalised}
\begin{align} 
        \cA^{(1)}_{\xx,\mathrm{ren}} = \ & \cA^{(1)}_{\xx,\mathrm{mren}}
         + \left( \frac{3}{2} \dZ_{\as}^{(1)} + \dZ_t^{(1)} \right) \cA^{(0)}_\xx \,, \\
        \cA^{(2)}_{\xx,\mathrm{ren}} = \ & A^{(2)}_{\xx,\mathrm{mren}} + \left(
        \frac{5}{2} \dZ_{\as}^{(1)} + \dZ_t^{(1)} \right) \cA^{(1)}_{\xx,\mathrm{mren}} \nonumber \\
        & + \left( \frac{3}{2} \dZ_{\as}^{(2)} +\frac{5}{2} \dZ_{\as}^{(1)}
        \dZ_{t}^{(1)} + \dZ_t^{(2)} + \frac{3}{8} (\dZ_{\as}^{(1)})^2
        \right) \cA^{(0)}_\xx \,.
\end{align}
\end{subequations}
The superscript $L$ in the renormalised partial amplitudes $\cA^{(L)}_{\xx,\mathrm{ren}}$, and generally for all fully renormalised quantities henceforth, refers to the order in the renormalised coupling, $\alpha_s/(4\pi)$.
The explicit expressions for $\dZ_{m}^{(L)}$, $\dZ_{t}^{(L)}$ and $\dZ_{\as}^{(L)}$ in the Feynman gauge are provided in \cref{app:renormalisation}.

Once UV singularities are removed, scattering amplitudes are still affected by divergences of IR origin. 
We define \emph{finite remainder} for each partial amplitude by subtracting the universal IR singularities~\cite{Catani:1998bh, 
Becher:2009cu, Becher:2009qa, Becher:2009kw, Gardi:2009qi, Gardi:2009zv, Catani:2000ef, Ferroglia:2009ep, Ferroglia:2009ii}
from the fully renormalised partial amplitude 
$\cA^{(L)}_{\xx,\mathrm{ren}}$, as
\begin{subequations}
\label{eq:finiteremainder}
\begin{align}
        \cF^{(0)}_\xx & = \cA^{(0)}_\xx \,, \\
        \cF^{(1)}_\xx & = \cA^{(1)}_{\xx,\mathrm{ren}} - \ZZ^{(1)}_\xx A^{(0)}_\xx \,, \\
	\cF^{(2)}_\xx & = \cA^{(2)}_{\xx,\mathrm{ren}} 
        - \ZZ^{(1)}_\xx \cA^{(1)}_{\xx,\mathrm{ren}} 
	- \left[ \ZZ^{(2)}_\xx - \big(\ZZ^{(1)}_\xx\big)^2 \right] \cA^{(0)}_\xx \,,
\end{align}
\end{subequations}
where the factor $\ZZ_x^{(i)}$ is the coefficient of $[\alpha_s/(4 \pi)]^i$ in the IR pole operator, $\ZZ_\xx$. 
In the $\overline{\mathrm{MS}}$ renormalisation scheme, $\ZZ_\xx$ is given by~\cite{Becher:2009cu,Becher:2009qa}
\begin{align} \label{eq:ZIR}
\begin{aligned}
	\ZZ_\xx = {} & 1+ \frac{\as}{4\pi} \left( \frac{\GammaPrime_{\xx,0}}{4\eps^2} + \frac{\GammaBold_{\xx,0}}{2\eps} \right) \,, \\
	&+  \left(\frac{\as}{4\pi}\right)^2 \left[ \frac{(\GammaPrime_{\xx,0})^2}{32\eps^4}
	+ \frac{\GammaPrime_{\xx,0}}{8\eps^3}\bigg(\GammaBold_{\xx,0} - \frac{3}{2} \beta_0 \bigg)
	+ \frac{\GammaBold_{\xx,0}}{8\eps^2} \big(\GammaBold_{\xx,0} - 2\beta_0 \big)
	+ \frac{\GammaPrime_{\xx,1}}{16\eps^2}
	+ \frac{\GammaBold_{\xx,1}}{4\eps} \right] \,.
\end{aligned}
\end{align}
The anomalous dimension coefficients $\GammaPrime_{\xx,n}$ and $\mathbf{\Gamma}_{\xx,n}$ are given at leading colour by
\begin{align} \label{eq:GammaNg}
\begin{aligned}
	\GammaPrime_{g,n} &= -3 \, N_c \gamma^{\mathrm{cusp}}_n \,, \\
	\GammaBold_{g,n} &= -N_c \frac{\gamma^{\mathrm{cusp}}_n}{2} \bigg(
        \mathcal{L}_{m,23} + \mathcal{L}_{m,15} + \mathcal{L}_{34} + \mathcal{L}_{45} \bigg)
        + 3\gamma^g_n+2 \gamma^Q_n \,,
\end{aligned}
\end{align}
for $\ttggg$, and
\begin{align} \label{eq:GammaNq}
\begin{aligned}
	\GammaPrime_{q,n} &= -2 \, N_c \gamma^{\mathrm{cusp}}_n \,, \\
	\GammaBold_{q,n} &= - N_c \frac{\gamma^{\mathrm{cusp}}_n}{2} \bigg(
                            \mathcal{L}_{m,23} + \mathcal{L}_{m,15} + \mathcal{L}_{45} \bigg)
                           + \gamma^g_n  +  2 \gamma^q_n  +  2 \gamma^Q_n \,,
\end{aligned}
\end{align}
for $\ttqqg$.
\Cref{eq:GammaNg,eq:GammaNq} involve the following logarithms of scalar invariants,\footnote{These logarithms are understood as analytically continued to the $s_{45}$ channel.}
\begin{align} \label{eq:Logs}
\begin{aligned}
        \mathcal{L}_{ij} &= -\log \bigl( -2d_{ij} \bigr) \,,  \\
        \mathcal{L}_{m,ij} &= -\frac{1}{2} \log \bigl( -2d_{ij} \bigr)
                               + \frac{1}{2} \log \biggl( \frac{m_t^2}{-2d_{ij}} \biggr) \,,
\end{aligned}
\end{align}
as well as the coefficients of the $\alpha_s$-expansion of the cusp and collinear anomalous dimensions, 
\begin{equation} \label{eq:gammaExpansion}
	\gamma^i \left(\alpha_s\right) = \sum_{n \geq 0} \gamma^i_n \bigg(\frac{\as}{4\pi}\bigg)^{n+1} \,,
\end{equation}
with $i \in \lbrace g,q,Q,\mathrm{cusp} \rbrace$.
The expressions of the anomalous dimension coefficients are given in \cref{app:renormalisation}.

The partial finite remainders $\cF^{(L)}$ have the same colour decomposition as the bare partial amplitudes
(c.f.~\cref{eq:colourdecompositionttggg,eq:colourdecompositionttqqg}), except that the bare strong coupling constant is replaced by its renormalised counterpart ($\alpha_s = g_s^2/(4 \pi)$). 
Therefore, the colour decomposition of the colour-dressed finite remainder $\cR^{(L)}$ in terms of the partial ones  $\cF^{(L)}$ is given by
\begin{align} \label{eq:colourdecompositionttgggfin}
\begin{aligned}
	\cR^{(L)}\left( 1_{\bar{t}}, 2_t, 3_g,4_g,5_g \right) =& \; \sqrt{2} \, g_s^3 \bigg[ (4\pi)^\eps e^{-\eps \gamma_E} \frac{\alpha_s}{4\pi} \bigg]^L   \\
	& \times \sum_{\sigma \in Z_3} \left( t^{a_{\sigma(3)}} t^{a_{\sigma(4)}} t^{a_{\sigma(5)}} \right)_{i_2}^{\;\bar{i}_1} \,
	\cF^{(L)}_{g}\left( 1_{\bar{t}}, 2_t, \sigma(3)_g, \sigma(4)_g, \sigma(5)_g \right)  \,,
\end{aligned}
\end{align}
for $\ttggg$, and
\begin{align} \label{eq:colourdecompositionttqqgfin}
\begin{aligned}
	\cR^{(L)}\left( 1_{\bar{t}}, 2_t, 3_{\qb},4_q,5_g \right) =& \; \sqrt{2} \, g_s^3 \bigg[ (4\pi)^\eps e^{-\eps \gamma_E} \frac{\alpha_s}{4\pi} \bigg]^L \times
          \bigg\lbrace \delta_{i_4}^{\;\bar{i}_1} \left( t^{a_5} \right)_{i_2}^{\;\bar{i}_3} \,
          \cF^{(L)}_{q}\left( 1_{\bar{t}}, 2_t, 3_{\qb}, 4_q, 5_g \right)  \\
        & \qquad + \delta_{i_2}^{\;\bar{i}_3} \left( t^{a_5} \right)_{i_4}^{\;\bar{i}_1} \,
	  \cF^{(L)}_{q}\left( 2_{\bar{t}}, 1_t, 4_{\qb}, 3_q, 5_g \right)  \bigg\rbrace \,,
\end{aligned}
\end{align}
for $\ttqqg$.
The partial finite reminders are further decomposed according to their $N_c$ and $n_f$ contributions as the bare partial amplitudes 
(c.f.~\cref{eq:NcNfdecomposition}):
\begin{subequations} \label{eq:FNcNfdecomposition}
\begin{align}
	\cF^{(1)}_{\xx} & = N_c \, F^{(1),N_c}_{\xx} + n_f \, F^{(1),n_f}_{\xx}  \,, \\
	\cF^{(2)}_{\xx} & = N_c^2 \, F^{(2),N_c^2}_{\xx} + N_c n_f \, F^{(2),N_c n_f}_{\xx} + n_f^2 \, F^{(2),n_f^2}_{\xx}  \,.
\end{align}
\end{subequations}

Finally, we employ the colour-dressed finite remainders to construct the colour- and polarisation-summed \textit{hard function} $\cH^{(L)}$, that will contribute to the cross section. 
Up to two loops, the hard function $\cH^{(L)}$ is defined as
\begin{subequations}
\label{eq:hardfunctions}%
\begin{align}
\cH^{(0)} & = \colhelsum \big|\cR^{(0)}\big|^2\,, \\
	\cH^{(1)} & = 2 \, \mathrm{Re} \, \colhelsum \cR^{(0)*} \cR^{(1)} \,, \\
\cH^{(2)} & = 2 \, \mathrm{Re} \,  \colhelsum \cR^{(0)*} \cR^{(2)}
             + \colhelsum \big|\cR^{(1)} \big|^2  \,,
\end{align}
\end{subequations}
where the overline indicates the average over colour and polarisation of initial states.
The computation of the hard functions in~\cref{eq:hardfunctions} requires the interference of amplitudes at different loop orders.
At leading colour, the colour-summed interference between $L_1$- and $L_2$-loop finite remainders is given in terms of partial finite remainders by
\begin{align}
\begin{aligned}
	& \sum_{\mathrm{colour}} \left[ \cR^{(L_1)}\left( 1_{\bar{t}}, 2_t, 3_g,4_g,5_g \right) \right]^* \cR^{(L_2)}\left( 1_{\bar{t}}, 2_t, 3_g,4_g,5_g \right) \\
	& \qquad\qquad = \; 2 \, g_s^6 \, \bigg[ (4\pi)^\eps e^{-\eps \gamma_E} \frac{\alpha_s}{4\pi} \bigg]^{L_1+L_2} \, \frac{N_c^4}{8} \\
	& \qquad\qquad\qquad \times  \sum_{\sigma \in Z_3} 
	   \left[\cF^{(L_1)}_{g}\left( 1_{\bar{t}}, 2_t, \sigma(3)_g, \sigma(4)_g, \sigma(5)_g \right) \right]^*
	   \cF^{(L_2)}_{g}\left( 1_{\bar{t}}, 2_t, \sigma(3)_g, \sigma(4)_g, \sigma(5)_g \right)  \,,
\end{aligned}
\end{align}
for $\ttggg$, while for $\ttqqg$ we have
\begin{align}
\begin{aligned}
	& \sum_{\mathrm{colour}} \left[ \cR^{(L_1)}\left( 1_{\bar{t}}, 2_t, 3_{\qb},4_q,5_g \right) \right]^* \cR^{(L_2)}\left( 1_{\bar{t}}, 2_t, 3_{\qb},4_q,5_g \right) \\
	& \qquad\qquad = \; 2 \, g_s^6 \, \bigg[ (4\pi)^\eps e^{-\eps \gamma_E} \frac{\alpha_s}{4\pi} \bigg]^{L_1+L_2} \, \frac{N_c^3}{2} \\
	& \qquad\qquad\qquad \times  \bigg\lbrace
	   \left[\cF^{(L_1)}_{q}\left( 1_{\bar{t}}, 2_t, 3_{\qb}, 4_q, 5_g \right) \right]^*
	   \cF^{(L_2)}_{q}\left( 1_{\bar{t}}, 2_t, 3_{\qb}, 4_q, 5_g \right) \\
	&  \qquad\qquad\qquad\qquad + \left[\cF^{(L_1)}_{q}\left( 2_{\bar{t}}, 1_t, 4_{\qb}, 3_q, 5_g \right) \right]^*
	     \cF^{(L_2)}_{q}\left( 2_{\bar{t}}, 1_t, 4_{\qb}, 3_q, 5_g \right) \bigg\rbrace \,.
\end{aligned}
\end{align}

\section{Analytic computation of the helicity amplitudes \label{sec:helamps}}

In this section we describe in detail the analytic computation of the leading colour $\ttggg$ and $\ttqqg$ amplitudes up to two-loop order in QCD.
We construct a representation of the amplitudes where the helicities of massless partons are fixed while the massive spinor dependence is written in terms of form factors.
We include the explicit dependence on the gluon and massless quark helicities $h_i$ in the process definitions in~\cref{eq:processdefinition} as
\begin{subequations}
\begin{align}
\label{eq:processhelicity}
	 & 0 \to \bar{t}(p_1) + t(p_2) + g(p_3,h_3) + g(p_4,h_4) + g(p_5,h_5)  \,, \\
	 & 0 \to \bar{t}(p_1) + t(p_2) + \bar{q}(p_3,h_3) + q(p_4,h_4) + g(p_5,h_5) \,.
\end{align}
\end{subequations}
The dependence on gluon and quark helicity states is treated within the spinor-helicity formalism, using the following conventions for massless quark wave functions and gluon polarisation vectors:
\begin{alignat}{3}
\begin{aligned}
& \bar{u}(p_i,+) = [i | \,, \qquad \qquad
& v(p_i,+) = | i ] \,, \qquad \qquad
& \vareps^\mu(p_i,p_j,+) = \frac{\spAB{j}{\gamma^\mu}{i}}{\sqrt{2} \, \spA{j}{i}}\,, \\
& \bar{u}(p_i,-) = \langle i | \,, \qquad \qquad
& v(p_i,-) = | i \rangle \,, \qquad \qquad
& \vareps^\mu(p_i,p_j,-)  = \frac{\spAB{i}{\gamma^\mu}{j}}{\sqrt{2} \, \spB{i}{j}}\,,
\end{aligned}
\end{alignat}
where $p_i$ and $p_j$ are massless momenta.
The second argument in the gluon polarisation vectors, $q_i$ in $\vareps^{\mu}(p_i,q_i)$, is an arbitrary reference momentum such that $q_i \cdot \vareps(p_i, q_i) = 0$;
we choose $q_3=p_4$, $q_4=p_3$, $q_5=p_3$ for $\ttggg$, and $q_5=p_3$ for $\ttqqg$.
The gluon polarisation vector sum is given by
\begin{equation} \label{eq:polsum}
	\sum_{h_i = \pm} \vareps(p_i,q_i, h_i)^{\mu \, *} \vareps(p_i,q_i, h_i)^{\nu} = -g^{\mu\nu} + \frac{ p^\mu_i q^\nu_i + q^\mu_i p^\nu_i  }{p_i \cdot q_i} \,.
\end{equation}

We decompose the amplitudes in terms of massive spinor form factors as~\cite{Buccioni:2023okz}
\begin{equation}
	\label{eq:massivespinordecomposition}
	\cA^{(L)h_3 h_4 h_5}_x = \sum_{i=1}^{4} \Psi_i \, \cG^{(L)h_3 h_4 h_5}_{x;i} \,,
\end{equation}
where
\begin{align}
\begin{aligned}
	\label{eq:massivespinortensorstructure}
	\Psi_1 & = m_t^2 \, \bar{u}(p_2) v(p_1) \,, \\
	\Psi_2 & = m_t \, \bar{u}(p_2) \slashed{p}_3 v(p_1) \,, \\
	\Psi_3 & = m_t \, \bar{u}(p_2) \slashed{p}_4 v(p_1) \,, \\
	\Psi_4 & =  \, \bar{u}(p_2) \slashed{p}_3  \slashed{p}_4 v(p_1) \,.
\end{aligned}
\end{align}
The form factors $\cG^{(L)h_3 h_4 h_5}_{x;i}$ are obtained via
\begin{equation}
	\cG^{(L)h_3 h_4 h_5}_{x;i} = \sum_{j=1}^{4} \left(\Omega^{-1}\right)_{ij} \; \tilde{\cA}^{(L)h_3 h_4 h_5}_{x;j}  \,,
\end{equation}
with
\begin{align}
	\Omega_{ij} & = \sum_{\mathrm{pol.}}\Psi^\dagger_i \Psi_j \,, \\
	\tilde{\cA}^{(L)h_3 h_4 h_5}_{x;i} & = \sum_{\mathrm{pol.}} \Psi^\dagger_i \,  \cA^{(L)h_3 h_4 h_5}_x \,,
\end{align}
where we sum over all massive spinor polarisations and the $\tilde{\cA}^{(L)h_3 h_4 h_5}_{x;i}$'s are referred to as \textit{contracted helicity amplitudes}.

In order to compute the contracted helicity amplitude we employ the physical projector method~\cite{Peraro:2019cjj,Peraro:2020sfm}. Before fixing the helicities of the massless partons, the amplitude can be decomposed into a set of tensor structures determined by the gluon and massless spinor polarisations, with the external momenta kept in four dimensions.
The tensor decomposition for $\ttggg$ and $\ttqqg$ are given~by
\begin{align}
	\label{eq:tensordecompositiongg}
	\tilde{\cA}^{(L)}_{g;i} &= \sum_{j=1}^{8} \cT_{g;j} \, \cS_{g;ij}^{(L)} \,, \\
	\label{eq:tensordecompositionqq}
	\tilde{\cA}^{(L)}_{q;i} &= \sum_{j=1}^{4} \cT_{q;j} \, \cS_{q;ij}^{(L)} \,,
\end{align}
where
\begin{align}
\label{eq:tensorstructurettggg}
\begin{aligned}
	\cT_{g;1} & = \vareps(p_3,q_3)\cdot p_1 \; \vareps(p_4,q_4)\cdot p_1 \; \vareps(p_5,q_5)\cdot p_1 \,, \\
	\cT_{g;2} & = \vareps(p_3,q_3)\cdot p_1 \; \vareps(p_4,q_4)\cdot p_1 \; \vareps(p_5,q_5)\cdot p_2 \,, \\
	\cT_{g;3} & = \vareps(p_3,q_3)\cdot p_1 \; \vareps(p_4,q_4)\cdot p_2 \; \vareps(p_5,q_5)\cdot p_1 \,, \\
	\cT_{g;4} & = \vareps(p_3,q_3)\cdot p_2 \; \vareps(p_4,q_4)\cdot p_1 \; \vareps(p_5,q_5)\cdot p_1 \,, \\
	\cT_{g;5} & = \vareps(p_3,q_3)\cdot p_1 \; \vareps(p_4,q_4)\cdot p_2 \; \vareps(p_5,q_5)\cdot p_2 \,, \\
	\cT_{g;6} & = \vareps(p_3,q_3)\cdot p_2 \; \vareps(p_4,q_4)\cdot p_1 \; \vareps(p_5,q_5)\cdot p_2 \,, \\
	\cT_{g;7} & = \vareps(p_3,q_3)\cdot p_2 \; \vareps(p_4,q_4)\cdot p_2 \; \vareps(p_5,q_5)\cdot p_1 \,, \\
	\cT_{g;8} & = \vareps(p_3,q_3)\cdot p_2 \; \vareps(p_4,q_4)\cdot p_2 \; \vareps(p_5,q_5)\cdot p_2 \,,
\end{aligned}
\end{align}
and
\begin{align}
\label{eq:tensorstructurettqqg}
\begin{aligned}
	\cT_{q;1} & = \bar{u}(p_4)\slashed{p}_1 v(p_3) \; \vareps(p_5,q_5)\cdot p_1 \,, \\
	\cT_{q;2} & = \bar{u}(p_4)\slashed{p}_1 v(p_3) \; \vareps(p_5,q_5)\cdot p_2 \,, \\
	\cT_{q;3} & = \bar{u}(p_4)\slashed{p}_2 v(p_3) \; \vareps(p_5,q_5)\cdot p_1 \,, \\
	\cT_{q;4} & = \bar{u}(p_4)\slashed{p}_2 v(p_3) \; \vareps(p_5,q_5)\cdot p_2 \,.
\end{aligned}
\end{align}
In \cref{eq:tensordecompositiongg,eq:tensordecompositionqq}, $\cS_{x;ij}^{(L)}$ are the form factors associated with the tensors in~\cref{eq:tensorstructurettggg,eq:tensorstructurettqqg}. They can be obtained similarly to $\cG^{(L)h_3 h_4 h_5}_{x;i}$ by means of
\begin{align}
	\cS_{g;ij}^{(L)} & = \sum_{k=1}^{8} \left(\Delta^{-1}_{g}\right)_{jk} \, \mathfrak{A}^{(L)}_{g;ki}  \,, \\
	\cS_{q;ij}^{(L)} & = \sum_{k=1}^{4} \left(\Delta^{-1}_{q}\right)_{jk} \, \mathfrak{A}^{(L)}_{q;ki}  \,,
\end{align}
where
\begin{align}
	\Delta_{x;ij} & = \sum_{\mathrm{pol}.} \cT_{x;i}^\dagger \cT_{x;j} \,, \\
	\mathfrak{A}^{(L)}_{x;ij} & = \sum_{\mathrm{pol}.} \cT_{x;i}^\dagger \, \tilde{\cA}^{(L)}_{x;j}
	                            = \sum_{\mathrm{pol}.} \cT_{x;i}^\dagger \, \Psi^\dagger_j \,  \cA^{(L)}_x   \,,
				    \label{eq:contractedamplitude}
\end{align}
where we sum over all spin polarisations.

In order to compute the contracted helicity amplitude $\tilde{\cA}^{(L)h_3 h_4 h_5}_{x;i}$, we fix the gluon and massless quark helicity states in
the tensor structures in~\cref{eq:tensordecompositiongg,eq:tensordecompositionqq} as follows:
\begin{subequations}
\begin{align}
	\tilde{\cA}^{(L)h_3 h_4 h_5}_{g;i} &= \sum_{j=1}^{8} \cT^{h_3 h_4 h_5 }_{g;j} \, \cS_{g;ij}^{(L)} \,, \\
	\tilde{\cA}^{(L)h_3 h_4 h_5}_{q;i} &= \sum_{j=1}^{4} \cT^{h_3 h_4 h_5} _{q;j} \, \cS_{q;ij}^{(L)} \,.
\end{align}
\end{subequations}
The explicit dependence of the tensor structures $\cT^{h_3 h_4 h_5}_{g;i}$ and $\cT^{h_3 h_4 h_5}_{q;i} $ on the gluon and massless quark helicities $h_3$, $h_4$ and $h_5$ is obtained by fixing the helicity of the gluon polarisation vectors and massless quark wave functions in \cref{eq:tensorstructurettggg,eq:tensorstructurettqqg}.

Finally, from the knowledge of $\tilde{\cA}^{(L)h_3 h_4 h_5}_{x;i}$, the \textit{contracted helicity finite remainders} $\tilde{\cF}^{(L)h_3h_4h_5}_{x;i}$ are obtained following \cref{eq:finiteremainder}, as
\begin{subequations}
\label{eq:helicityfiniteremainder}
\begin{align}
	\tilde{\cF}^{(0)h_3h_4h_5}_{\xx;i} & = \tilde{\cA}^{(0)h_3h_4h_5}_{\xx;i} \,, \\
	\tilde{\cF}^{(1)h_3h_4h_5}_{\xx;i} & = \tilde{\cA}^{(1)h_3h_4h_5}_{\xx;i,\mathrm{ren}} - \ZZ^{(1)}_\xx \tilde{A}^{(0)h_3h_4h_5}_{\xx;i} \,, \\
	\tilde{\cF}^{(2)h_3h_4h_5}_{\xx;i} & = \tilde{\cA}^{(2)h_3h_4h_5}_{\xx;i,\mathrm{ren}} 
	- \ZZ^{(1)}_\xx \tilde{\cA}^{(1)h_3h_4h_5}_{\xx,\mathrm{ren}} 
	- \left[ \ZZ^{(2)}_{\xx;i} - \big(\ZZ^{(1)}_\xx\big)^2 \right] \tilde{\cA}^{(0)h_3h_4h_5}_{\xx;i} \,.
\end{align}
\end{subequations}
We note that the tensor decompositions of the partial amplitudes in \cref{eq:massivespinortensorstructure,eq:tensordecompositiongg,eq:tensordecompositionqq}
equally apply to the partial finite remainders. Specifically, we compute the following set of independent helicity configurations:
\begin{subequations}
\label{eq:independenthelicities}
\begin{align}
	\ttggg & : \qquad \tilde{\cF}^{(L)+++}_{g;j},\quad \tilde{\cF}^{(L)++-}_{g;j},\quad \tilde{\cF}^{(L)+-+}_{g;j}\,,  \\
	\ttqqg & : \qquad \tilde{\cF}^{(L)+-+}_{q;j},\quad \tilde{\cF}^{(L)+--}_{q;j}\,. 
\end{align}
\end{subequations}
The remaining helicity configurations can be obtained by parity conjugation and/or permutation of external momenta.
For stable top-quarks in the final state, the interference between $L_1$- and $L_2$-loop helicity finite remainders required for the hard functions is given in terms of the contracted helicity finite remainders by
\begin{equation}
	\sum_{\mathrm{pol}.} \left[\cF_x^{(L_1)h_3 h_4 h_5}\right]^* \cF_x^{(L_2)h_3 h_4 h_5}
	= \sum_{i,j=1}^{4} \left[\tilde{\cF}_{x;i}^{(L_1)h_3 h_4 h_5}\right]^* \, \left(\Omega^{-1}\right)_{ij} \, \tilde{\cF}_{x;j}^{(L_2)h_3 h_4 h_5} \,,
\end{equation}
where we sum over the massive top polarisations.
Our representation for the amplitude makes the inclusion of top-quark decays within the narrow-width approximation straightforward: the decay amplitude can be directly attached to the massive spinor structures in \cref{eq:massivespinortensorstructure}~\cite{Melnikov:2009dn,Campbell:2012uf}.

We compute analytically the contracted helicity finite remainders $\tilde{\cF}^{(L)h_3 h_4 h_5}_{x;i}$ for the independent helicity configurations with the same framework successfully employed in a number of two-loop five-point amplitude calculations~\cite{Badger:2022ncb,Badger:2024sqv,Badger:2024mir},
combining Feynman diagrams, four-dimensional projectors, integration-by-part (IBP)
reduction, special function bases and finite-field arithmetic.
We construct the loop integrand by generating the Feynman diagrams for the colour structures contributing 
to $\ttggg$ and $\ttqqg$ in the leading colour approximation using \textsc{Qgraf}~\cite{Nogueira:1991ex}.
The resulting amplitudes are then contracted by the conjugated tensor structures in
\cref{eq:massivespinortensorstructure,eq:tensorstructurettggg,eq:tensorstructurettqqg} to obtain the unreduced contracted
amplitude $\mathfrak{A}^{(L)}_{x;ij}$ according to \cref{eq:contractedamplitude} by means of in-house \textsc{Form}~\cite{Ruijl:2017dtg}
and \textsc{Mathematica} scripts to identify the loop integral families, perform the Dirac algebra
and simplify expressions.
At this stage, the contracted amplitudes $\mathfrak{A}^{(L)}_{x;ij}$ are written as a linear combination
of scalar Feynman integrals. We subsequently express the scalar integrals present in the amplitude in terms of 
the master integrals of refs.~\cite{Badger:2022hno,Badger:2024fgb} by IBP reduction~\cite{Tkachov:1981wb,Chetyrkin:1981qh,Laporta:2000dsw}.
We generate compact systems of IBP relations with \textsc{NeatIBP}
package~\cite{Wu:2023upw} and solve them numerically over finite fields with \textsc{FiniteFlow}'s sparse linear solver~\cite{Peraro:2016wsq,Peraro:2019svx}.
We then write the master integrals as polynomials in a set of special functions---denoted cumulatively by $\vec{f}$---and transcendental constants order by order in $\eps$ as discussed in \cref{sec:specialfunction}.
This enables the analytic subtraction of UV and IR poles according to \cref{eq:finiteremainder}.
To this end, we also compute the one-loop amplitudes up to order $\eps^2$ with the same framework.
The resulting finite remainders are expressed in terms of monomials of special functions and transcendental constants $m_k(\vec{f} \,)$ as
\begin{equation}
	\tilde{\cF}_{\xx;i}^{(L) h_3 h_4 h_5} = \sum_{k} r_{\xx;i,k}^{(L) h_3 h_4 h_5} \; m_k(\vec{f} \,) \,.
	\label{eq:finiteremainderspfunc}
\end{equation}

We obtain the analytic expression of the coefficients $r_{\xx;i,k}^{(L) h_3 h_4 h_5}$ in \cref{eq:finiteremainderspfunc} from numerical finite-field evaluations.
Starting from the unreduced contracted amplitudes $\mathfrak{A}^{(L)}_{\xx;ij}$, known analytically, all subsequent
rational operations to derive $r_{\xx;i,k}^{(L)}$ are done numerically with finite-field arithmetic and concatenated in a dataflow graph within the
\textsc{FiniteFlow} framework~\cite{Peraro:2016wsq,Peraro:2019svx};
these operations include the solution of the IBP relations,
the map of master integrals onto the special functions, the subtraction of UV/IR poles as well as the matrix operations required to derive the contracted helicity
finite remainders $\tilde{\cF}^{(L)h_3 h_4 h_5}_{\xx;i}$.
In order to enable the use of finite-field arithmetic in the helicity amplitude computation, we employ the momentum-twistor parametrisation introduced in~\cite{Badger:2022mrb}.
As a result, the coefficients $r_{\xx;i,k}^{(L)}$ in \cref{eq:finiteremainderspfunc} are rational functions of the following momentum-twistor variables,
\begin{equation}
	\label{eq:momentumtwistorvariables}
	\vec{t} = (s_{34},t_{12},t_{23},t_{45},t_{51},x_{5123}) \,,
\end{equation}
related to the invariants in \cref{eq:dijmtset,eq:tr5definition} as
\begin{align}
	\begin{aligned}
		s_{34} & = 2 d_{34} \,, \qquad t_{12} = \frac{d_{12}+m_t^2}{d_{34}} \,, \qquad t_{23} = \frac{d_{23}}{d_{34}} \,, \qquad
		 t_{45}  = \frac{d_{45}}{d_{34}} \,, \qquad\; t_{51} = \frac{d_{15}}{d_{34}} \,,   \\
	 x_{5123}  & = \frac{4 \left[ d_{12} d_{23} - d_{23} d_{34} + d_{34} d_{45} - d_{45} d_{15} + d_{15} d_{12} + m_t^2 (d_{23} + d_{15}) \right] + \mathrm{tr}_{5}}{8 (d_{12} + m_t^2) (d_{34} + d_{45} - d_{12} - m_t^2)} \,.
\end{aligned}
\end{align}
For the same reason, we remove all square roots from the definition of the master integrals of refs.~\cite{Badger:2022hno,Badger:2024fgb}, effectively moving them from the coefficients $r_{\xx;i,k}^{(L)}$---that would otherwise be algebraic---to the special function monomials $m_k(\vec{f} \,)$ in \cref{eq:finiteremainderspfunc}.
We omit the dependence of the monomials on the square roots to simplify the notation.
Note that $s_{34}$ is the only dimensionful variable in our momentum-twistor parametrisation, hence it can be set to $1$ to simplify the computation and reintroduced by dimensional analysis.
In order to permute or conjugate momentum-twistor expressions, it is necessary to eliminate their spinor phases through division by suitable phase factors.
We use the following:
\begin{align}
	\Phi_g^{+++} = \frac{\spB{3}{5}\spB{3}{4}}{\spA{3}{5}}\,, \qquad \quad
	\Phi_g^{++-} = \frac{ \spAA{5}{3}{4}{5} }{ \spA{3}{4}^2 }  \,, \qquad \quad
	\Phi_g^{+-+} = \frac{ \spAA{4}{5}{3}{4} }{ \spA{3}{5}^2 }  \,,
\end{align}
for $\ttggg$ and
\begin{align}
	\Phi_q^{+-+} = \frac{ \spA{3}{4}\spB{3}{5} }{ \spA{3}{5} } \,, \qquad \quad
	\Phi_q^{+--} = \frac{ \spA{4}{5}\spB{3}{4} }{ \spB{4}{5} } \,,
\end{align}
for $\ttqqg$.
We refer to appendix~C of ref.~\cite{Badger:2023mgf} for a thorough discussion of how to permute and conjugate momentum-twistor expressions.

We employ the functional reconstruction strategy proposed in refs.~\cite{Badger:2021nhg,Badger:2021imn}
to reconstruct the analytic expression of the rational coefficients $r_{\xx;i,k}^{(L) h_3 h_4 h_5}$ in \cref{eq:finiteremainderspfunc} from numerical finite-field evaluations.
We collect in \cref{tab:reconstructiondata} some data regarding the reconstruction to illustrate the complexity of this computation and the effectiveness of our reconstruction algorithm.
We focus on the most complicated partial finite remainders: those for the process $\ttggg$ 
at order $N_c^2 $ ($\tilde{\cF}^{(2),N_c^2}_{g;i}$).
We include simultaneously all the independent helicity configurations and set $s_{34}$ to 1.
As a proxy for the degree of algebraic complexity, we show the maximum total polynomial degree of numerators and denominators, that is linked to how many evaluations are required to complete the reconstruction.
To demonstrate the advantage of expressing master integrals in a special function basis, we first give the degree data of the mass-renormalised contracted helicity partial amplitudes (see \cref{eq:massrenormalised}) in the master-integral representation;
in this case, the degrees refer to the rational coefficients of the master integrals---rather than the special function monomials---and include the full dependence on the regulator $\eps$.
We observe that our special function basis not only allows for the analytic subtraction of UV and IR poles, but also decreases significantly the maximum degrees.
The subsequent rows give the maximum degrees at various stages of our reconstruction strategy: searching for and solving linear relations among the coefficients, matching the denominator factors from the knowledge of potential singularities (derived from the differential equations for the master integrals, see \cref{sec:specialfunction})~\cite{Abreu:2018zmy},
and partial fraction decomposition in one kinematic variable (which we choose to be $x_{5123}$).
We refer the reader to section~4 of ref.~\cite{Badger:2021imn} for an in-depth discussion of each optimisation stage employed in the analytic reconstruction, as listed in the first column of \cref{tab:reconstructiondata}.
Finally, we need to reconstruct analytically the numerators of the coefficients of the univariate partial fraction decomposition;
to this end, we employ \textsc{FiniteFlow}'s built-in reconstruction algorithm and the last row of \cref{tab:reconstructiondata} gives the number of required sample points.\footnote{
The coefficients we reconstruct are functions of the momentum-twistor variables $(t_{12},t_{23},t_{45},t_{51})$ and by ``sample point'' we mean a numerical value for this set of variables.
Each sample point therefore entails the evaluation at several values of $\eps$ and $x_{1523}$, as required to extract the coefficients of the Laurent expansion around $\eps=0$ and of the univariate partial fraction decomposition in $x_{1523}$. 
See ref.~\cite{Badger:2021imn} for a thorough discussion.
}
To lift the constants from finite fields to rational numbers, the numerical evaluation needs to be performed in 2 prime fields for each sample point; this reflects the complexity of the appearing rational numbers
and is in line with what is observed in other two-loop five-particle amplitude computations.
The evaluation time for each sample point quoted in the last row of \cref{tab:reconstructiondata} is $\approx 25$ minutes (measured on \textit{Intel(R) Xeon(R) Gold 6342 CPU @ 2.80GHz}) and
the evaluation of all sample points needed to complete the functional reconstruction of the two-loop leading colour amplitudes
for both $\ttggg$ and $\ttqqg$ took around four months with $\cO(8)$ multicore machines (with 48--64 threads used on each machine, requiring $\approx 200$ GB of RAM).
Thanks to the univariate partial fraction decomposition performed to optimise the reconstruction, the resulting analytic expressions are by construction simplified and can be directly implemented in a \text{C++} library, as discussed in \cref{sec:res}.

\renewcommand{\arraystretch}{1.5}
\begin{table}[]
\centering
\begin{tabular}{|c|c|c|c|c|}
\hline
	& $1$ & $2$ & $3$ & $4$ \\
\hline
\hline
	\makecell{master integral coefficients \\ of mass-renormalised amplitude \\ (full $\eps$ dependence)} & 404/393 & 398/389 & 411/402 & 421/411 \\
\hline
	\makecell{special function coefficients \\ of finite remainder}           & 314/303 & 305/296 & 321/312 & 326/317 \\
\hline
	linear relations                                             & 291/280 & 287/278 & 299/293 & 304/299 \\
\hline
	denominator matching \#1                                     & 291/0   & 287/0   & 299/0   & 304/0   \\
\hline
	\makecell{partial fraction decomposition \\ in $x_{5123}$}   & 44/40   & 55/51   & 57/54   & 58/54   \\
\hline
	denominator matching  \#2                                    & 44/0    & 54/0    & 54/0    & 56/0    \\
\hline
\hline
	\makecell{number of sample points \\ (1 prime field)}        & 137076  & 89624   & 161482  & 179838  \\
\hline
\end{tabular}
\caption{Maximum total numerator/denominator polynomial degrees of the rational coefficients appearing in the most complicated $\ttggg$ partial mass-renormalised amplitudes and finite remainders ($N_c^2$-components, see \cref{eq:NcNfdecomposition}). 
The first row refers to the four components of the massive spinor decomposition in \cref{eq:massivespinortensorstructure}.
The degree data for the finite remainders are shown at different stages of the optimisation strategy of refs.~\cite{Badger:2021nhg,Badger:2021imn}.
All the independent helicity configurations are evaluated simultaneously, $\vec{h} \in \lbrace +++,++-,+-+ \rbrace $, and $s_{34} =1$.
The last row gives the number of sample points required to complete the functional reconstruction in a prime field.}
\label{tab:reconstructiondata}
\end{table}


\section{A full set of special functions and their numerical evaluation \label{sec:specfuncs}}
\label{sec:specialfunction}

\subsection*{Basis of functions and differential equations}

In this section, we extend the representation of master integrals in terms of special functions constructed in ref.~\cite{Badger:2024dxo} to all crossings of the external momenta relevant to the cross section.
Our guiding principle is to write the required terms of the Laurent expansion of the MIs around $\eps=0$ as polynomials in a set of algebraically independent special functions to resolve all redundancies, enabling simplification and analytic cancellation of UV/IR poles in the amplitudes.
The presence of elliptic functions in a subset of the MIs~\cite{Badger:2024fgb} prevents us from following the established procedure~\cite{Gehrmann:2018yef,Chicherin:2020oor,Chicherin:2021dyp,Abreu:2023rco,Gehrmann:2024tds}, that relies on canonical DEs for the MIs~\cite{Henn:2013pwa}.
While canonical DEs are now available also for the MIs involving elliptic functions~\cite{Becchetti:2025oyb}, the numerical evaluation of their solution remains an open problem.
Therefore, we continue with the hybrid approach proposed in ref.~\cite{Badger:2024dxo}.
By working with algebraic, non-canonical DEs, this method isolates the terms of the MIs' $\eps$-expansion that do not satisfy canonical DEs into few (potentially related) functions that only contribute to the finite part of the two-loop amplitude, while treating the remaining terms with the established algorithm to obtain algebraically independent functions.

The MI representation of ref.~\cite{Badger:2024dxo} covers only the permutations of the Feynman integral families that contribute to the helicity partial amplitudes (see fig.~2 there), hereby referred to as the ``minimal'' set of permutations.
Instead, all 12 permutations of the external momenta that preserve the on-shell conditions in \cref{eq:onshell} contribute to the hard function in \cref{eq:hardfunctions}.
The additional permutations can be included in the same formalism straightforwardly with the algorithm presented in sec.~4 of ref.~\cite{Badger:2024dxo}.
The starting point are the DEs for the MIs obtained in refs.~\cite{Badger:2024fgb} for a single ordering of the external momenta, written in terms of $\d\log$ (alias letters) and non-$\d\log$ one-forms.
By permuting these, we obtain DEs for all needed permutations of the MIs, introducing new square roots, letters and non-$\d\log$ one-forms.
We choose a spanning basis of $\mathbb{Q}$-linearly independent one-forms, preferring those with the lowest maximal polynomial degrees.
The number of independent square roots and one-forms and a comparison against the subset considered in ref.~\cite{Badger:2024dxo} are given in  \cref{tab:alphabet}.
We then used \textsc{AMFlow}~\cite{Liu:2017jxz,Liu:2022chg} (interfaced to \textsc{LiteRed}~\cite{Lee:2013mka} and \textsc{FiniteFlow}~\cite{Peraro:2019svx}) to evaluate numerically all MIs with (at least) 60-digit precision in three random points in the physical $s_{45}$ channel.
Finally, we employed the algorithm of ref.~\cite{Badger:2024dxo} to expand the MIs around $\eps=0$ up to $\eps^4$ and express the expansion terms as polynomials with constant rational coefficients in
\begin{itemize}
 \item a set of algebraically independent pure transcendental functions $\{f_i^{(w)}(\vec{d}\,)\}$ with transcendental weight $w$ up to $4$ (see ref.~\cite{Henn:2013pwa} for the notions of transcendental weight and pure functions);
 \item a set of functions $\{f_i^{(4^*)}(\vec{d}\,)\}$ that cover the MI expansion terms associated with nested square roots and elliptic curves, whose DEs are not in canonical form;
 \item a set of algebraically independent transcendental constants ($\zeta_2$ and $\zeta_3$).
\end{itemize}
We call the set of special functions $\vec{f} \equiv \{f_i^{(w)}, \forall \, w =1,\ldots,4,4^*, \forall \, i\}$ \emph{generating set}.
The number of functions and a comparison against the subset of permutations considered in ref.~\cite{Badger:2024dxo} are shown in \cref{tab:functions}.
The special functions in the generating set inherit from the MIs they are defined from an even or odd parity with respect to flipping the sign of each square root.
Unlike the MIs, instead, the special functions are not homogeneous functions of the invariants $\vec{d}$.
This follows by construction from the fact that the MIs have an $\eps$-dependent scaling dimension, and that the special functions are defined from the terms of the $\eps$-expanded MIs.
As a result of this and of the fact that we set $\mu_R^2 = 1$, the finite remainders in \cref{eq:finiteremainderspfunc} do not have a uniform scaling dimension, although the rational coefficients
$r_{\xx;i,k}^{(L) h_3 h_4 h_5}$ do.
This inhomogeneity of the special functions is compensated by logarithms of the renormalisation scale $\mu_R$ in the full amplitudes and finite remainders, so that the latter are homogeneous in $s_{34}$ (the only dimensionful variable in our momentum-twistor parametrisation) and $\mu_R^2$ as dictated by dimensional analysis.
The dependence of the finite remainders on logarithms of $\mu_R$ is restored as discussed in \cref{app:mudep}.

\begin{table}[h]
\begin{center}
\begin{tabular}{c||c|c|c|c|c}
 & square roots & rat.~letters & alg.~letters & all letters & non-$\d \log$ one-forms \\
\hline
minimal & 9 & 51 & 58 & 109 & 126 \\
\hline
all & 26 & 121 & 144 & 265 & 566 \\
\hline
all (up to $\epsilon^4$) & 26 & 109 & 132 & 241 & 338 \\
\end{tabular}
\end{center}
\caption{Number of independent square roots, letters (split into rational and algebraic) and non-$\d \log$ one-forms appearing in the 1- and 2-loop Feynman integrals at leading colour.
``Minimal'' refers to the permutations considered in ref.~\cite{Badger:2024dxo};
``all'' includes all 12 permutations that preserve the on-shell conditions in \cref{eq:onshell};
``up to $\eps^4$'' means that the MIs are truncated at order $\eps^4$, thus keeping only what is relevant to the two-loop amplitudes up to order $\eps^0$.
}
\label{tab:alphabet}
\end{table}

\begin{table}[h]
\begin{center}
\begin{tabular}{c|c||c|c|c|c|c|c|c}
\multicolumn{2}{c||}{weight} & $0$ & $1$ & $2$ & $3$ & $4$ & $4^*$ & all \\
\hline
\multirow{2}{*}{minimal} & generating set & $0$ & 6 & 8 & 45 & 166 & 12 & 237\\
& DE basis & $1$ & 6 & 30 & 151 & 166 & 12 & 366 \\
\hline
\multirow{2}{*}{all} & generating set & $0$ & 11 & 22 & 167 & 699 & 48 & 947 \\
& DE basis & $1$ & 11 & 80 & 443 & 699 & 48 &  1282 \\
\end{tabular}
\end{center}
\caption{Number of special functions, of each weight and total, in our representation of the MIs (``generating set'') and in the linear system of first-order DEs (see \cref{eq:eqF}) we use for the numerical evaluation (``DE basis'').
For simplicity, we list $4^*$ in the row for the transcendental weight, even though these functions do not have pure transcendentality.
The sets ``minimal''/``all'' are defined in \cref{tab:alphabet}.
The DE basis includes also monomials of constants of weight 0, 2 and 3 ($1$, $\zeta_2$, $\zeta_3$).}
\label{tab:functions}
\end{table}

While we do not have complete analytic control over the functional relations satisfied by the MI expansion terms involving nested square roots and elliptic functions, we identified integer relations among them from their numerical values with the PSLQ algorithm~\cite{Ferguson1998APT}, and used these to reduce the number of functions in the set $\{f_i^{(4^*)}\}$.
The resulting relations coincide with those derived from the symmetry arguments below.
Fixing the ordering of the external legs to the one considered in ref.~\cite{Badger:2024fgb}, there are 6 MIs of family ${\rm PB}_{B}$ whose expansion terms at order $\eps^4$ involve non-$\d\log$ one-forms: 4 are associated with the elliptic curve (MIs \# $15$, $35$, $36$, $37$), and 2  with the nested square roots (MIs \# $19$ and $20$).
The polynomial defining the elliptic curve (eq.~(A.6) of ref.~\cite{Badger:2024fgb}) is invariant with respect to swapping two of the massless momenta; this implies that only 6 out of 12 permutations give different elliptic curves, leading to $6 \times 4$ functions.
We verified that the 6 elliptic curves are not isomorphic to each other by computing their $j$-invariants.
The nested square roots instead depend on the full five-particle kinematics (eq.~(3.25) of ref.~\cite{Badger:2024fgb}), resulting in $12 \times 2$  functions.
Together, we have 48 functions $f_i^{(4^*)}$'s, defined from suitable permutations of the order-$\eps^4$ terms of the MIs \# $\{15,19,20,35,36,37\}$ of ${\rm PB}_{B}$, whose DEs are not in canonical form and involve non-$\d\log$ one-forms.
Applying the PSLQ algorithm to numerical values allowed us not only to check the trivial relations stemming from symmetries but also to exclude the existence of further non-trivial $\mathbb{Q}$-linear relations.
We emphasise that the $f_i^{(4^*)}$'s only appear in the MIs at order $\eps^4$ (equivalently, at order $\eps^0$ in the two-loop amplitude), hence missing relations involving them would not prevent the analytic cancellation of UV/IR poles in the two-loop amplitude.

The generating set of special functions is by construction closed under the 12 relevant permutations of the external momenta.
In other words, any permutation of any $f_i^{(w)}(\vec{d} \,)$ can be expressed as a polynomial with constant rational coefficients in the transcendental constants ($\zeta_2$, $\zeta_3$) and the generating set of special functions themselves, evaluated at the same phase-space point $\vec{d}$.
The explicit relations can be found by expressing $f_i^{(w)}(\vec{d} \,)$ in terms of MI expansion terms, applying the permutation to the MIs, and writing the result back in the generating set of special functions.
As a result, evaluating the generating set of special functions at a single phase-space point in the $s_{45}$ channel suffices to evaluate all permutations of the amplitudes that contribute to the cross section.
This allows us to avoid analytic continuation to other channels, and reduces the number of required evaluations of special functions by resolving all redundancies among the permutations.
We refer to sec.~2.5 of ref.~\cite{Badger:2023mgf} for a thorough discussion of this approach.

In order to evaluate numerically the special functions, we construct a linear system of first-order DEs for them, derived from the DEs for the MIs~\cite{Badger:2021nhg}.
In general, the derivatives of a special function contain products of special functions.
Therefore, we need to complement the $f_i^{(w)}$'s by a minimal set of polynomials in the $f_i^{(w)}$'s themselves, $\zeta_2$ and $\zeta_3$ (including the constant, $1$) in order to close them under derivatives.
We follow the procedure outlined in sec.~4.2 of ref.~\cite{Badger:2024dxo}.
We call the resulting set of 1282 functions \emph{DE basis}, denote it by $G(\vec{d} \,)$, and break it down in \cref{tab:functions}.
The DEs for $G(\vec{d} \,)$ have the form
\begin{align} \label{eq:eqF}
 \d G(\vec{d} \,) =  M(\vec{d} \,) \cdot G(\vec{d} \,) \,,
\end{align}
where $\d$ is the total differential in the kinematic variables $\vec{d}$.
The connection matrix $M(\vec{d} \,)$ is given by a linear combination of constant rational matrices $M_i$ and $\mathbb{Q}$-linearly independent differential one-forms $\omega_i(\vec{d} \, )$, as
\begin{align} \label{eq:eqFM}
M(\vec{d} \,) = \sum_{i} M_i \, \omega_i( \vec{d} \, ) \,.
\end{align}
Here, the one-forms are a subset of those relevant for the MIs, as we truncate the latter at order $\eps^4$;
the relevant numbers are given in the last row of \cref{tab:alphabet}.
The $\d\log$ one-forms have the form
\begin{align}
 \omega_i(\vec{d} \,) = \d \log W_i(\vec{d}\,) \,,
\end{align}
where $W_i(\vec{d}\,)$ is an algebraic function called a letter in the literature~\cite{Henn:2013pwa},
while we write the non-$\d\log$ one-forms as
\begin{align} \label{eq:omega2dx}
 \omega_j(\vec{d} \,) = \sum_{x \in \vec{d}} \omega_{j,x}(\vec{d} \,) \, \d x \,,
\end{align}
where $\omega_{j,x}(\vec{d} \,)$ is an algebraic function.
We provide explicit the expression of the letters, the non-$\d\log$ one-forms, the connection matrix and all MIs in terms of the generating set of special functions in our ancillary files~\cite{zenodo}.
Each one-form is either even or odd with respect to flipping the sign of each square root, and square roots are factored out overall.
Note that, while the DEs in \eqref{eq:eqF} satisfy integrability conditions, most of the non-$\d\log$ one-forms are not closed;
care must thus be taken that all one-forms are parametrised on the same path when integrating between two endpoints.
The system is sparse ($M(\vec{d} \,)$ has roughly $3 \%$ of non-zero entries) and has the following block structure,
\begin{align}
G(\vec{d} \, ) = \begin{pmatrix}
                            f^{(4^*)}_i \\
                            f^{(4)}_i \\
                            \text{weight-3} \\
                            \text{weight-2} \\
                            f^{(1)}_i \\
                            1 \\
                           \end{pmatrix} \,,
\qquad \qquad
M(\vec{d} \, ) = \begin{pmatrix}
							 Y_{4^*,4^*} & 0 & Y_{4^*,3} & Y_{4^*,2} & 0 & 0 \\
							 0 & 0 & X_{4,3} & 0 & 0 & 0 \\
							 0 & 0 & 0 & X_{3,2} & 0 & 0 \\
							 0 & 0 & 0 & 0 & X_{2,1} & 0 \\
							 0 & 0 & 0 & 0 & 0 & X_{1,0} \\
							 0 & 0 & 0 & 0 & 0 & 0 \\
                            \end{pmatrix} \,,
\end{align}
where ``weight-$w$'' in $G(\vec{d} \, )$ denotes uniform weight-$w$ polynomials in the special functions and the transcendental constants,
$X_{w,w-1}$ in $M(\vec{d} \, )$ denotes a non-zero block containing only $\d\log$ one-forms, while the non-zero blocks $Y_{4^*,w}$ also involve non-$\d\log$ one-forms.
This structure follows from the fact that the derivatives of pure weight-$w$ functions contain only weight-$(w-1)$ functions, whereas the functions $f^{(4^*)}_i$ originate from MIs whose DEs are quadratic in $\eps$~\cite{Badger:2024fgb}.

The efficiency of the numerical solution of DEs is strongly correlated with the smoothness of the derivatives of the basis functions. Special functions are an improvement with respect to MIs because many singularities disappear, as can be seen by comparing the last two rows of \cref{tab:alphabet}. Furthermore, the connection matrix $M(\vec{d} \,)$ is substantially more sparse than for a set of MIs and the system is by construction free of linear redundancies.
The same holds also for the representation of the connection matrix in \cref{eq:eqFM}, where the basis of one-forms is chosen to be as simple as possible. These properties reduce the evaluation time of the derivatives at every step of the numerical solution. We achieve a further substantial speed-up by the following set of manipulations aimed at reducing the number of floating-point operations.

\begin{enumerate}
 \item We set $m_t^2 = 1$.
Unlike MIs and amplitudes, our special functions $f_i^{(w)}$ are not homogeneous functions of $\vec{d}$ because the renormalisation scale is set to unity. The functions are evaluated for rescaled kinematics $\vec{d}/m_t^2$, while the scale dependence is restored at the end according to appendix~\ref{app:mudep} for the renormalization scale $\mu_R/m_t$. The correct value of the hard functions \eqref{eq:hardfunctions} is obtained by multiplication with $1/m_t^2$ as dictated by dimensional analysis.

 \item We tailor the parametrisation of the kinematics to each one-form $\omega_i(\vec{d} \,)$ in \cref{eq:eqFM}.
 We scan over all changes of variables $\vec{d} \to \vec{d}'$ obtained by permuting the momenta in the definition of the invariants $\vec{d}$, and choose the one that yields the smallest number of multiplications in the factored form of $\omega_i(\vec{d}' \,)$. The change of variables includes the differentials on the right-hand side of \cref{eq:omega2dx}.
 For instance, we write $\omega_{140}$ as a function of $\vec{d}' = \sigma_{21453} \circ \vec{d}$, where the permutation acts on the momenta as $\sigma_{21453} \circ (p_1,p_2,p_3,p_4,p_5) = (p_2,p_1,p_4,p_5,p_3)$.
In terms of the original variables, we have that $\vec{d}' = \bigl(d_{12}, d_{23} - d_{45} - d_{15} , d_{45}, 1 + d_{12} - d_{34} - d_{45}, d_{23}, 1\bigr)$. For $\omega_{140}$, this change of variables reduces the number of multiplications by $\approx 65 \%$.
A similar procedure to change variables was employed in ref.~\cite{Badger:2022ncb} to simplify the amplitude's rational coefficients.

\item We decompose into multivariate partial fractions the coefficients of the differentials in each one-form ($\omega_{j,x}(\vec{d} \,)$ in \cref{eq:omega2dx}) with \textsc{MultivariateApart}~\cite{Heller:2021qkz} interfaced to \textsc{Singular}~\cite{DGPS} for the computation of Gr\"obner bases (square roots are factored overall).

\item We factor each partial fraction, and write all one-forms in terms of a global set of multiplicatively independent polynomials.

\item We optimise the expression of the polynomials with \textsc{Mathematica}'s function \texttt{Simplify} to reduce the number of floating-point operations.

\end{enumerate}
This optimisation reduces the connection-matrix file size by about an order of magnitude.

\subsection*{Numerical evaluation of the special functions}

At this point, we have an optimised set of DEs together with high-precision values $\vec{f}_{0i} \equiv \vec{f}(\vec{d}_{0i})$ at a few random $\vec{d}_{0i}$ points. The special functions $\vec{f}$ can now be evaluated at any given phase-space point $\vec{d}$ by solving the differential-equation system numerically \cite{Boughezal:2007ny,Czakon:2008zk} along a curve $\gamma \equiv \{ \vec{d}_{\gamma}(t) \}_{t \in [0,1]}$, with $\vec{d}_{\gamma}(0) = \vec{d}_{0i}$ for a chosen $i$ and $\vec{d}_{\gamma}(1) = \vec{d}$. In principle, the top-quark mass is fixed at $m_t = 1$, but the methodology is valid for a variable top-quark mass as well. For the numerical solution of the DEs, we use the Bulirsch-Stoer algorithm (see for example \cite{Hairer}) implemented in the \textsc{OdeInt} library of \textsc{Boost} \cite{BoostLibrary} as done first for this type of problems in \cite{Czakon:2020vql,Czakon:2021yub}. Let us now describe the main aspects of our implementation.

\begin{enumerate}

\item The DEs are solved for the 947 independent special functions. Hence, {\bf the system is non-linear}, while the derivatives of the functions are still given by \eqref{eq:eqF}, where only the relevant rows of the connection matrix are kept. There is no reason to solve for all the 1282 functions which include linear combinations of the 947 independent ones, because the Bulirsch-Stoer algorithm does not profit from a linear system. Unfortunately, we do not gain anything noticeable in performance, because all the one-forms are still present in the 947 rows of the connection matrix.

\item The curve $\gamma$ is a {\bf complex deformation} of a curve ${\gamma}_{\text{R}} \subset \mathbb{R}^6$,
  \begin{equation}
    d_{\gamma,j}(t) = d_{0i,j} + \big( 1 + 4 (1-t) \, \mathrm{i} \, \delta_j \big) \big( d_{\gamma_\text{R},j}(t) - d_{0i,j} \big) \; , \qquad j = 1,\dots,6 \; .
  \end{equation}
  This deformation is necessary, since the connection matrix is singular on subsets of the phase space to the extent that it is impossible to leave the vicinity of $\vec{d}_{0i}$ using finite-precision numerics despite the smoothness of the special functions. We only deform the positive invariants $d_{12}$, $d_{23}$ and $d_{45}$. This choice is based on the physicist's prejudice that positive invariants include an infinitesimal imaginary part. In the present case, no imaginary parts are required for the correctness of the results as the curve ${\gamma_\text{R}}$ is chosen to lie in the physical phase space above threshold. The imaginary parts are only used to achieve numerical stability. Hence, any of the invariants could have been deformed. The values of the parameters $\delta_j$ should neither be excessively large nor excessively small. We set the deformation parameters by default to $0.1$, $0.15$ and $0.2$ for the three invariants respectively. It turns out that the values should be chosen all different, otherwise the solver hits numerical instabilities for some quite ordinary phase-space points.

\item The curve ${\gamma_\text{R}}$ is chosen to be a straight line,
  \begin{equation}
    \vec{d}_{\gamma_\text{R}}(t) = \vec{d}_{0i} + t \big( \vec{d} - \vec{d}_{0i} \big) \; ,
  \end{equation}
  if it is entirely {\bf contained in the physical phase space}. In principle, this requirement could be relaxed with additional care to end up on the correct Riemann sheet of the special functions at $t = 1$. We avoid the pitfalls of analytic continuation by switching to a different curve if necessary. This curve is linearly parameterised by $t$ in $d_{12}$, $d_{45}$, $m_t^2$, the cosine of the gluon scattering angle in the center-of-mass frame, as well as the cosines of the polar and azimuthal angle of the top quark in the rest frame of the $t\bar{t}$ pair. This curve is obviously longer than a straight line. We have also used it for points which can be reached by a straight line, but the computational time is worsened at the level of a few percent on average. It should be pointed out, however, that it is not the length of the curve that is a reliable indicator for the efficiency of the numerical integration. Rather, it is the behaviour of the special functions along the integration curve in the complex space that impacts the number of evolution steps and hence also the total integration time. It seems that an optimal curve cannot be determined a priori. Nevertheless, the performance achieved is satisfactory, and we have not studied any other curves.

\item The DEs contain square roots while some of the special functions are odd with respect to a change of the branch of a single square root. 
The choice of branch of each square root is arbitrary and does not affect the amplitude, but must be handled consistently. We choose the standard principal branch, adopted by the standard system libraries, i.e.\ square roots are either real positive or have positive imaginary part.
The square roots remain on the same branch along ${\gamma_\text{R}}$. The complex curve ${\gamma}$, however, is defined without guarantee to remain on a single branch of all the square roots. For this reason, we have implemented the following {\bf algorithm that yields smooth square roots along the integration curve}. Before calling the DE solver, the software determines brackets, $[t_a,t_b] \subset [0,1]$, for the integration variable $t$, that contain a single branch-cut crossing for some square root. During integration, the software verifies whether the integration variable $t$ is within a bracket. If this is the case, it uses the signs of the imaginary part of the relevant square root at the ends of the bracket and at $t$ to determine the correct branch, which might result in a sign change of the square root. After integration, the sign of an odd function is changed if the relevant square root crosses the branch cut an odd number of times. In order to determine the brackets, the integration curve is divided into a number of points (1000 by default) uniformly in $t$, and a search is performed for pairs of points for which the real part of the argument of a square root is negative while the imaginary part changes sign. Due to the finite graining of the division, branch-cut crossings might be missed. In order to guarantee correctness of the values of $\vec{f}$, the smoothness of the square roots is analysed after integration, which is made possible by storing the values of the square roots for each sampled $t$. In particular, the software searches for sign changes of the imaginary part of the square roots which do not agree with expected values obtained by interpolation. This check cannot be done on the fly, because $t$ is not necessarily increased with each call. If it is discovered that some square root is not smooth, the complete procedure is restarted with 10 times more sampling points for the bracket search. Notice that the search for brackets is substantially cheaper than a search for actual branch-cut crossings. Furthermore, the procedure can be generalised to nested square roots and even arbitrary analytic functions.

We should point out that the original differential-equation system for the master integrals derived directly from integration-by-parts identities does not contain square roots. We have compared the timings for the solution of a set of DEs without square roots for the 237 special functions that do not take into account permutations. The solution was about 20\% slower than for the corresponding system with square roots.  

\item The Bulirsch-Stoer algorithm requires local error estimates in order to determine the size of the evolution step. The \textsc{Boost} library provides a default error checker, which uses error estimates for each special function to determine a total error, which can be a combination of absolute and relative errors and can even take into account the size of the derivatives of the special functions. We are, however, not interested in a precise determination of the special functions, but rather in a {\bf precise determination of the amplitude}. In order to achieve a predefined precision, we have modified the error checker as follows. The normalised amplitude for $g_s = 1$ derived from \eqref{eq:hardfunctions},
  \begin{equation}\label{eq:func_vec}
    \mathcal{Q} \equiv \frac{1}{(4\pi)^4} \frac{2 \,  \colhelsum \cR^{(0)*} \cR^{(2)}}{\colhelsum \big|\cR^{(0)}\big|^2} \; ,
  \end{equation}
  is computed as a vector of 3380 coefficients of monomials in the special functions. For given estimated values $\vec{f}(t)$ and errors $\Delta \vec{f}(t)$, we set the error estimate of the normalised amplitude to be
  \begin{equation} \label{eq:errorchecker}
    \Delta \mathcal{Q}(t) = \bigg[ \sum_{i=1}^{947}  \Big| \frac{\partial Q}{\partial f_i} \Big|_{\vec{f} = \vec{f}(t)}^2 \, \big| \Delta f_i(t) \big|^2 \bigg]^{1/2} \; .
  \end{equation}
  It turns out that 24 weight-4 non-elliptic special functions do not contribute to the $gg \to t\bar{t}g$ amplitude. For simplicity of implementation, we nevertheless still evaluate the complete set which is needed for the amplitudes with light quarks on external lines. The irrelevant functions simply do not contribute to the error estimate. We observe that requesting a given precision for $\Delta \mathcal{Q}(t)$ \eqref{eq:errorchecker} yields a result for $Q(t = 1)$ (see next point) with an error of the same order. The Bulirsch-Stoer algorithm is thus very robust in the sense that local and global precision are nearly the same. As far as implementation is concerned, we had to modify the \textsc{Boost} code, because, contrary to other DE solvers in the same library, the error checker is hardcoded. Fortunately, such a modification is allowed by the copyright.

\item The {\bf error of the amplitude} is determined by performing the integration starting from two different $\vec{d}_{0i}$ points and comparing the results. The initial points are chosen at random, and we observe that one of them always requires noticeably more evaluation steps. Nevertheless, since the overall performance is satisfactory, we have not tried to optimise the choice of the initial points w.r.t.\ the endpoint of the integration. In practice, double precision is sufficient for about 10 digits precision at most phase-space points. If required, the integrals can be evaluated with quadruple or even higher precision. The limiting factor is the precision of the initial values, which are currently provided with 60 digits.

\end{enumerate}

\noindent
The performance of the numerical integration is illustrated in \cref{tab:performance}.

\begin{table}[h]
\begin{center}
\begin{tabular}{c||c|c|c|c|c|c|c}
  \makecell{float type \\ boundary} &  \makecell{requested \\ $|\Delta Q|$} & \makecell{achieved \\ $|\Delta Q|$} & \makecell{maximal \\ $|\Delta f_i/f_i|$} & \makecell{number \\ of steps} & \makecell{number \\ of calls} & \makecell{call \\ time} & \makecell{time} \\
  \hline \hline
  \makecell{\texttt{double} \\ $\vec{d}_{01}$} & $10^{-10}$ & $8.2 \times 10^{-10}$ & $1.4 \times 10^{-6}$ & 17 & 716 & 0.77 ms & 0.56 s \\
  \hline
  \makecell{\texttt{double} \\ $\vec{d}_{02}$} & $10^{-10}$ & $8.2 \times 10^{-10}$ & $1.4 \times 10^{-6}$ & 28 & 1184 & 0.73 ms & 0.89 s \\
  \hline
  \makecell{\texttt{dd\_real} \\ $\vec{d}_{01}$} & $10^{-10}$ & $7.7 \times 10^{-11}$ & $1.3 \times 10^{-6}$ & 16 & 709 & 9.3 ms & 6.7 s \\
  \hline
  \makecell{\texttt{dd\_real} \\ $\vec{d}_{02}$} & $10^{-10}$ & $7.7 \times 10^{-11}$ & $1.3 \times 10^{-6}$ & 27 & 1166 & 9.3 ms & 11 s \\
  \hline
  \makecell{\texttt{dd\_real} \\ $\vec{d}_{01}$} & $10^{-20}$ & $3.8 \times 10^{-21}$ & $1.2 \times 10^{-16}$ & 42 & 3002 & 9.0 ms & 27 s \\
  \hline
  \makecell{\texttt{dd\_real} \\ $\vec{d}_{02}$} & $10^{-20}$ & $3.8 \times 10^{-21}$ & $1.2 \times 10^{-16}$ & 68 & 4873 & 9.0 ms & 44 s \\
\hline
\end{tabular}
\end{center}
\caption{Performance benchmarks for the numerical integration of DEs on SUSE Linux Enterprise 15, Intel(R) Core(TM) i9-10980XE CPU @ 3.00GHz, software compiled with \texttt{g++-13 -O2}. The numbers are averages over 100 points corresponding to unweighted Born events in the $gg \to t\bar{t}g$ channel. Achieved $|\Delta Q|$ and maximal $|\Delta f_i/f_i|$ have been determined by comparing the results obtained starting from two different points $\vec{d}_{01}$ and $\vec{d}_{02}$. The floating-point types used are either the standard \texttt{double} or \texttt{dd\_real} from the \textsc{QD} library~\cite{QD}. Number of calls and call time refer to calls to the function that evaluates the connection matrix. The achieved $|\Delta Q|$ in the first two rows is dominated by a single point that only has $7.0 \times 10^{-8}$ precision.}
\label{tab:performance}
\end{table}

Before closing this section, let us comment of the recent publication~\cite{PetitRosas:2025xhm}, which contains a reappraisal of the numerical solution of DEs for high multiplicity problems. The authors take two two-loop integral families from the leading colour $gg \to t\bar{t} g$ amplitude as one of their examples. They achieve a similar performance to ours, if we restrict our system to the same integral set. Although the Bulirsch-Stoer algorithm from \textsc{Boost} is used in both cases, there are important differences between our approach and the approach presented in ref.~\cite{PetitRosas:2025xhm}. In particular, a linear system of DEs is integrated one variable at a time along a complex curve in ref.~\cite{PetitRosas:2025xhm} just as in ref.~\cite{Czakon:2021yub}. Furthermore, the integration curve for each variable is chosen so that the square roots remain on the same branch, which is achieved by factorising the arguments of the square roots into irreducible polynomials. The publication provides a strategy to optimise the length of the integration curve made of multiple segments. Error control is based on the standard error checker, and the authors discuss how to estimate the total error after integration along all of the segments. While integration along one smooth curve, as we do, has a certain elegance to it, there is one potential advantage to integrating in each variable separately. Indeed, the connection matrix can be precomputed as a function of just one single variable with the remaining variables replaced by their numerical values. The evaluation time of the connection matrix is then vastly reduced and the numerical stability improved. At the present moment, it is not possible to judge which strategy is better in general. This will be an interesting question once more variables are involved in a higher-multiplicity problem. 


\section{Results \label{sec:res}}

The analytic expression of the rational coefficients discussed in \cref{sec:helamps} and the evaluation strategy of the special functions presented in \cref{sec:specialfunction} allow us to assemble an efficient implementation of the two-loop leading colour $pp\to t\bar{t}j$ amplitude that is suitable for direct phenomenological applications.
In our ancillary files~\cite{zenodo}, we provide the analytic expression of the contracted helicity finite remainders $\tilde{F}^{(L) h_3 h_4 h_5}_{x;i}$ and scripts to evaluate numerically the mass-renormalised contracted 
helicity amplitudes $\tilde{A}^{(L) h_3 h_4 h_5}_{\mren,x;i}$  and hard functions $\cH^{(L)}$ in \textsc{Wolfram}'s \textsc{Mathematica} format, as well as a \text{C++} library for fast numerical evaluation.

\subsection*{Checks}
We have validated our results as follows:
\begin{itemize}
	\item The analytic pole cancellation necessary to derive the finite remainders proves that our results are consistent with the universal structure of UV/IR singularities (see \cref{sec:partial}).

	\item We have checked the dependence of the finite remainders on the renormalisation scale ($\mu_R$),
		as provided in~\cref{app:mudep}, by comparing the numerical values of the finite remainders at a random phase-space point $\vec{p}$ with $\mu_R = Q_0 \ne 1$ against
		an evaluation at the rescaled phase-space point $\vec{p}/Q_0$ with $\mu_R=1$, and verifying that the ratio of the two results is as prescribed by dimensional analysis.

	\item We have reproduced the numerical benchmarks provided in ref.~\cite{Badger:2024dxo} for $\ttggg$. To this end, we translated the massive spinor structures in \cref{eq:massivespinortensorstructure} into the notation of refs.~\cite{Badger:2022mrb,Badger:2024dxo} to link the contracted helicity amplitudes derived
		in this work to the the helicity subamplitudes $A^{(L),[i]}$ in eq.~(3.4) of ref.~\cite{Badger:2024dxo}.

	\item We cross checked the one-loop amplitude (up to $\cO(\eps^2)$) required for the two-loop UV and IR counterterms against ref.~\cite{Bera:2025upg}.
	We note the full-colour one-loop $pp\to t\bar{t}j$ amplitudes derived in ref.~\cite{Bera:2025upg}
		have been thoroughly cross checked against \textsc{OpenLoops}~\cite{Buccioni:2019sur} up to $\cO(\eps^0)$.
		The comparison against the leading colour one-loop amplitudes derived in this paper is achieved by removing
		the sub-leading colour contributions in the full-colour amplitude of ref.~\cite{Bera:2025upg}.
\end{itemize}

\subsection*{Benchmarks}
We will now provide benchmark numerical values for the hard functions defined in \cref{eq:hardfunctions}.
We give numerical results for all scattering channels relevant for the $t\bar{t}j$ production with stable top quarks; these include those from the $\ttggg$ partonic process,
\begin{align}
        gg\to \tb t g &: \quad g(-p_4)+g(-p_5) \to \tb(p_1) + t(p_2) + g(p_3) \,,
\end{align}
and those from $\ttqqg$,
\begin{align}
\begin{aligned}
        q\qb  \to \tb t g &: \quad q(-p_4)+\qb(-p_5) \to \tb(p_1) + t(p_2) + g(p_3) \,, \\
        \qb q \to \tb t g &: \quad \qb(-p_4)+ q(-p_5) \to \tb(p_1) + t(p_2) + g(p_3) \,, \\
        q g \to \tb t q &: \quad q(-p_4)+ g(-p_5) \to \tb(p_1) + t(p_2) + q(p_3) \,, \\
        g q \to \tb t q &: \quad g(-p_4)+ q(-p_5) \to \tb(p_1) + t(p_2) + q(p_3) \,, \\
        \qb g \to \tb t \qb &: \quad \qb(-p_4)+ g(-p_5) \to \tb(p_1) + t(p_2) + \qb(p_3) \,, \\
        g \qb \to \tb t \qb &: \quad g(-p_4)+ \qb(-p_5) \to \tb(p_1) + t(p_2) + \qb(p_3) \,.
\end{aligned}
\end{align}
In \cref{tab:num-hardfunction} we present the numerical values of the hard functions evaluated
at a random physical phase-space point in the $s_{45}$ channel. 
We choose the same benchmark point as in ref.~\cite{Badger:2024fgb}:
\begin{alignat}{3}
\begin{aligned}
    &d_{12}=\frac{4602}{57095} \, \mathrm{GeV}^2 \,, \qquad
    && d_{23}=\frac{217}{8151} \, \mathrm{GeV}^2 \,, \qquad
    &&d_{34}=-\frac{8513}{67193} \, \mathrm{GeV}^2 \,,  \\
    & d_{45}=\frac{7}{22} \, \mathrm{GeV}^2 \,, && d_{15}=-\frac{14291}{77626} \, \mathrm{GeV}^2 \,, && m_t^2  = \frac{1701}{90164} \, \mathrm{GeV}^2 \,.
\end{aligned}
\label{eq:pspoint}
\end{alignat}
The strong coupling constant and the renormalisation scale are chosen to be
\begin{equation}
	\alpha_s = 1 \,, \qquad \qquad \mu_R = 17 \, \mathrm{GeV} \,. 
\label{eq:parameter}
\end{equation}

\renewcommand{\arraystretch}{1.5}
\begin{table}
    \centering
    \begin{tabular}{|c|ccc|}
    \hline
    $\ttggg$ & $\cH^{(0)}$ [GeV$^{-2}$] & $\cH^{(1)}/\cH^{(0)}$ & $\cH^{(2)}/\cH^{(0)}$  \\
    \hline
    $gg \to \bar{t}tg$ & 268350.649723 & $-$26.4682819001  & 322.779928392  \\
    \hline
    \hline
    $\ttqqg$ & $\cH^{(0)}$ [GeV$^{-2}$] & $\cH^{(1)}/\cH^{(0)}$ & $\cH^{(2)}/\cH^{(0)}$  \\
    \hline
    $q\bar{q} \to \bar{t}tg$       &  84846.1219815  & $-$12.2457197210  &  77.1742474127  \\
    $\bar{q}q \to \bar{t}tg$       &  39186.3029738  & $-$15.0696085114  &  109.314968526  \\
    $qg \to \bar{t}tq$             &  63115.3293016  & $-$7.62559522058  &  50.4934637358  \\
    $gq \to \bar{t}tq$             &  88243.7106496  & $-$11.5275362050  &  80.4225454869  \\
    $\bar{q}g \to \bar{t}t\bar{q}$ &  30252.0699036  & $-$11.5393226140  &  79.2838889381  \\
    $g\bar{q} \to \bar{t}t\bar{q}$ &  44510.2654758  & $-$9.21990859071  &  58.9236118458  \\
    \hline
    \end{tabular}
	\caption{Benchmark numerical results for the $pp\to t\bar{t}j$ hard functions (defined in~\cref{eq:hardfunctions}) in the leading colour approximation at the physical point and parameters given 
	in~\cref{eq:pspoint,eq:parameter}.
	}
    \label{tab:num-hardfunction}
\end{table}
Numerical results for the mass-renormalised contracted helicity amplitude are provided in the ancillary files for the independent colour structures
($ \left( t^{a_{3}} t^{a_{4}} t^{a_{5}} \right)_{i_2}^{\;\bar{i}_1}$ for $\ttggg$ and 
$\delta_{i_4}^{\;\bar{i}_1} \left( t^{a_5} \right)_{i_2}^{\;\bar{i}_3}$ for $\ttqqg$), 
independent helicity configurations (c.f.\ \cref{eq:independenthelicities}) and unpermuted scattering channels 
($gg\to \tb t g$ and $\qb g \to \tb t \qb$) with $\mu_R=1$~GeV.

\subsection*{C++ implementation}\label{subsec:cpp}
In addition to the analytical expressions, a numerical implementation of the squared matrix elements is provided~\cite{zenodo}.
Besides example programs described below, the \text{C++} package consists of two parts: the implementation of the differential equation solver for the special functions as described in~\cref{sec:specialfunction} and a code that evaluates the rational coefficients and assembly of the squared matrix elements as defined in~\cref{eq:hardfunctions}.
The code requires a specification of the partonic channel, a set of four momenta $\{p_i\}_{i=1,5}$, the top-quark mass $m_t$, and the renormalisation scale $\mu_R$ as input (the couplings are set to $g_s=1$ or equivalently $\alpha_s = 1/(4\pi)$).

The evaluation proceeds as follows to leverage the error-control strategy outlined in~\cref{sec:specialfunction}.
First, the input momenta and renormalisation scale are rescaled by $m_t$.
Then the 3484 rational coefficients in front of the special functions are evaluated (this includes all required permutations as well as the purely rational, $\zeta_2$ and $\zeta_3$ terms which increases the number of elements with respect to \cref{eq:func_vec}).
The coefficients are implemented as sparse matrices and analytic expressions, which are loaded and parsed at runtime.
The polynomial expressions are stored in a custom text format that represents a Horner form.
The coefficients are then used as weights for the numerical solution of the special functions as described in \cref{sec:specialfunction} and to perform the error control on the squared matrix element level.
If the test on the numerical precision of the special functions is successful, we evaluate the 3484 special function monomials ($m_k(\vec{f})$ in \cref{eq:finiteremainderspfunc}, including $\zeta_2$ and $\zeta_3$)---not required for the accuracy test in \cref{eq:errorchecker}---and combine them with the rational coefficients to obtain the helicity amplitudes.
The $\mu_R$-depending parts are restored using the equation in~\cref{app:mudep} for $\mu'_R = \mu_R /m_t$ and the squared matrix elements are assembled by performing the colour and helicity sums.
After restoring the overall $m_t$ rescaling of the amplitudes, the results for the hard functions in~\cref{eq:hardfunctions} are returned.

The code uses higher floating-point precision through the \textsc{QD} package~\cite{QD}.
In particular, the evaluation of the rational functions proved numerically challenging, to the extent that evaluation using standard \texttt{double} floating-point numbers is not sufficiently stable to yield meaningful values of the squared two-loop hard functions.
For that reason, the default is to use the \texttt{dd\_real} type for the rational coefficients and \texttt{double} for the special functions.
A checked evaluation of the amplitude using a rescaling test is also provided.

The package has been tested to reproduce the number in~\cref{tab:num-hardfunction} (this test can be repeated using the \texttt{validate} program provided) and has been used to successfully evaluate about 90k physical top-quark pair plus jet phase-space points at the 13 TeV LHC.
The average evaluation times broken down into coefficients, special functions, and assembly, are shown in~\cref{tab:ampcpp_num}.
Most of the time is spent evaluating the rational functions and their permutations: about 30 seconds for the $gg$ channel and 10 seconds for the $qg$ and $q\bar{q}$ channels.
The evaluation of the special functions for two boundary points and the assembly takes about 3--4 seconds in total.
The test can be repeated using the \texttt{performance} program.

\begin{table}
\centering
\begin{tabular}{l|c|c|c|c}
Channel & Functions [s]& Coefficents [s]& Assembly [s]& total [s]\\
\hline
$gg\to \bar{t}tg$            & 2.69 & 30.90 & 1.58 & 35.17 \\
$\bar{q}g\to \bar{t}t\bar{q}$& 2.16 &  9.40 & 0.18 & 11.74 \\
$qg\to \bar{t}tq$            & 2.50 &  9.62 & 0.21 & 12.33 \\
$q\bar{q}\to \bar{t}tg$      & 2.12 &  9.30 & 0.18 & 11.60\\
\hline
\end{tabular}
\caption{Evaluation times in seconds for the different channels including a break down into different stages. The numbers have been obtained with on Ubuntu 22.04, with a 12th Gen Intel(R) Core(TM) i5-1240P processor, and compilation with \texttt{gcc-11 -O2}. The difference in timing for the special functions in comparison with~\cref{tab:performance} is due to the evaluation for two boundary points and a different CPU. A substantial variance with respect to the used CPU and compilation options has been observed and should be kept in mind when comparing to these numbers, implying that they are for indicative purposes only.}
\label{tab:ampcpp_num}
\end{table}

\begin{figure}
\centering
\includegraphics[width=0.5\textwidth]{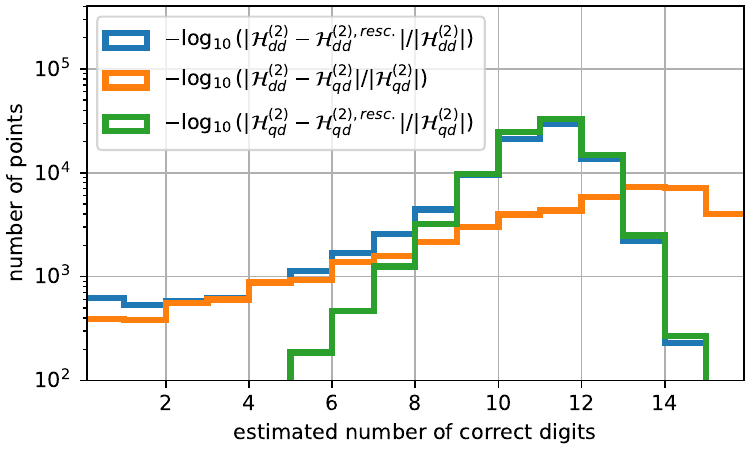}
\caption{Estimated number of correct digits of the two-loop hard function for evaluations with \texttt{dd\_real}, in blue, and \texttt{qd\_real}, in green, floating point precision for the rational coefficients. The number of correct digits is estimated through a scaling test. A comparison to the relative difference between \texttt{dd\_real} and \texttt{qd\_real} evaluations is shown in orange. The special functions are evaluated in \texttt{double} precision in both cases.}
\label{fig:stability}
\end{figure}

The numerical stability of the hard function has been tested on a set of phase-space points by using the implemented scaling test and a comparisons between evaluations using \texttt{dd\_real} and \texttt{qd\_real} floating-point types for the rational coefficients.
For the scaling test, all momenta, the top-quark mass and the renormalisation scale have been multiplied by a factor $f$ (here $f=10$) and the resulting amplitude multiplied with the expected energy scaling ($f^2$).
The number of agreeing digits between the original and scaled amplitude is shown in~\cref{fig:stability} for both floating point types. Additionally, the numerical stability has been gauged by comparing the results for \texttt{dd\_real} and \texttt{qd\_real} floating-point types directly.
The good agreement between the scaling test for \texttt{dd\_real} and the comparison to the more stable \texttt{qd\_real} evaluations for unstable phase space points (i.e. below approximately 5 correct digits) indicates the reliability of the scaling test and error control on the special functions as outlined in section \cref{sec:specialfunction}.
To achieve 5-digit accuracy on the two-loop hard function, we needed about 4\% to be evaluated with \texttt{qd\_real} precision.

Finally, we give some technical remarks on the usage of the provided package.
The memory load can be quite high, driven by the rational coefficients, and at least 16 GB of RAM is recommended.
The first evaluation during runtime takes substantially (3--4 times) longer than the given average, since the analytic expressions need to be read and interpreted; afterwards, they are in memory, and the quoted evaluation times can be achieved.
We only implement four of the channels ($gg\to \bar{t}tg$, $\bar{q}g\to \bar{t}t\bar{q}$, $qg\to \bar{t}tq$, and $q\bar{q}\to \bar{t}tg$) explicitly since all others can be obtained by permuting the input momenta.
A program to directly evaluate the special functions, dubbed \texttt{integrate}, is provided.
More details on the usage of the provided libraries and programs can be found in the software documentation.

\section{Conclusions}

In this article we have completed the computation of the two-loop double
virtual QCD corrections to the process $pp\to t\bar{t} j$ in the leading colour
approximation. This provides the final ingredient required for the production
of differential cross section predictions at NNLO in QCD, in the aforementioned
approximation, necessary to keep theoretical uncertainties inline with current
experiments. Modern techniques were essential to overcome bottlenecks at many
stages of the computation owing to the considerable analytic and algebraic
complexity. Using four-dimensional projectors, we reconstructed analytic forms
for the two-loop helicity finite remainders from finite field evaluations. The
use of univariate partial fractioning in the momentum-twistor variables was
essential to reduce the polynomial degrees to a manageable level. Another
substantial challenge is the appearance of elliptic structures in the master
integrals and their impact on the computation. In particular, the numerical
evaluation of these functions in a phenomenologically viable way remains an
open problem. We addressed this issue by constructing a (possibly)
over-complete basis of special functions that satisfy DEs involving only
algebraic functions~\cite{Badger:2024dxo}. Furthermore, we presented a new
method for the numerical solution of the differential equations for the special
functions making use of the Bulirsch-Stoer algorithm that shows promising
performance in terms of both speed and numerical stability also in view of
general applications to other complicated Feynman integrals. We implemented the
evaluation of both the special functions and their rational coefficients in the
amplitudes into a \texttt{C++} library suitable for
phenomenological~studies~\cite{Badger:2025ilt}. We note that the representation of the finite
remainders presented here can also be used to include top quark decays in the
narrow width approximation.

The final results do raise some open questions. In our implementation we
see that the size of the rational coefficients and large cancellations between
terms in the hard function are the bottleneck in the computation and often
require quadruple precision numerics. It would certainly be beneficial to
explore possibilities to reduce this cost in future implementations, in
particular when looking at direct comparison with experimental data and the
extraction of SM parameters, which put additional strain on Monte Carlo
simulations. In fact, recent presentations of the two-loop amplitudes for
$pp\to 3j$~\cite{DeLaurentis:2023nss,DeLaurentis:2023izi} and $pp\to
Vjj$~\cite{DeLaurentis:2025dxw}, building on the approach proposed in
ref.~\cite{DeLaurentis:2022otd}, show that such optimisations are not
without hope. Applications of this procedure with massive fermions have also been possible at
one-loop~\cite{Campbell:2025ftx}. There have even been proposals that Machine
Learning algorithms could be used to simplify spinor-helicity
expressions~\cite{Cheung:2024svk}, although there would need to be further
developments before applications to the expressions derived here could be
attempted. Lifting the leading colour approximation is desirable, indeed all
other contributions at NNLO are known at full colour, but will introduce
another level of complexity which will require new ideas to overcome. 

\section*{Acknowledgements}
S.B.\ and C.B.\ acknowledge funding from the Italian Ministry of Universities and Research (MUR) through FARE grant R207777C4R and through grant PRIN 2022BCXSW9.
M.B.~acknowledges funding from the European Union’s Horizon Europe research and innovation programme under the ERC Starting Grant No.~101040760 \emph{FFHiggsTop}. M.C.\
 acknowledges funding by the Deutsche Forschungsgemeinschaft (DFG) under grant 396021762-TRR 257: Particle Physics Phenomenology after the Higgs Discovery.
H.B.H.~has been supported by an appointment to the JRG Program at the APCTP through the Science and Technology Promotion Fund and Lottery Fund of the Korean Government and by the Korean Local Governments~--~Gyeongsangbuk-do Province and Pohang City.
S.Z.~was supported by the Swiss National Science Foundation (SNSF) under the Ambizione grant No.~215960. The authors gratefully acknowledge the computing time provided
to them at the NHR Center NHR4CES at RWTH Aachen University (project number p0020025). R.P. acknowledges that this research was funded in part by NCN 2024/55/D/ST2/00934.

\appendix

\section{Renormalisation factors and anomalous dimensions}
\label{app:renormalisation}
In this appendix we list the renormalisation constants entering the subtraction of UV singularities according to \cref{eq:massrenormalised,eq:UVrenormalised}. Their expansion 
in the strong coupling constant is
\begin{align}
	\bar\alpha_s &= \alpha_s \; \bigg\lbrace 1 + \bigg(\frac{\alpha_s}{4\pi}\bigg) \dZ_{\alpha_s}^{(1)} 
	+ \bigg(\frac{\alpha_s}{4\pi}\bigg)^2 \dZ_{\alpha_s}^{(2)} \bigg\rbrace + \cO\left( \alpha_s^4 \right) \,, \\
	Z_{m} &= 1 + \bigg(\frac{\bar\alpha_s}{4\pi}\bigg) \dZ_{m}^{(1)} + \bigg(\frac{\bar\alpha_s}{4\pi}\bigg)^2 \dZ_{m}^{(2)} + \cO\left( \bar\alpha_s^3 \right) \,, \\
	Z_{t} &= 1 + \bigg(\frac{\bar\alpha_s}{4\pi}\bigg) \dZ_{t}^{(1)} + \bigg(\frac{\bar\alpha_s}{4\pi}\bigg)^2 \dZ_{t}^{(2)} + \cO\left( \bar\alpha_s^3 \right) \,, 
\end{align}
where $\bar\alpha_s$ and $\alpha_s$ are the bare and renormalised coupling constants, respectively. The mass-renormalisation counterterms  are provided in the integral representation
to allow for the direct computation of the mass-renormalised amplitude in our finite-field framework. 
In the tHV scheme, keeping only terms contributing at leading colour, they are given by~\cite{Badger:2021owl}
\begin{align} \label{eq:dZm}
	\dZ_m^{(1)} = {} & N_c \; \frac{(1-\eps)(3-2\eps) }{2 (1-2\eps) m_t^2} \; I_1 \,, \\
        \dZ_m^{(2)} = {} & N_c^2 \; \bigg\lbrace
	  - \frac{(1-\eps)^2 (3-4\eps^2) }{4 (1-2\eps)^2 m_t^4}  \, I_2
	  + \frac{(1-\eps) ( -7 + 8\eps + 12\eps^2-16 \eps^3) }{4(1-2\eps)(1-4\eps)m_t^2} \, I_3 \bigg\rbrace \nonumber \\
	  & + N_c n_f \, \frac{(1-\eps) (1-2\eps) }{(1-4\eps) m_t^2} \, I_3 \,.
\end{align}
The integrals $I_i$ are defined by
\begin{align} \label{eq:dZmInt}
        I_1 &= e^{\eps \gamma_E} \int \dk{1} \; \frac{1}{k_1^2-m_t^2} \,, \\
        I_2 &= e^{2\eps \gamma_E} \int \dk{1}\dk{2} \; \frac{1}{(k_1^2-m_t^2) (k_2^2-m_t^2)} \,, \\
        I_3 &= e^{2\eps \gamma_E} \int \dk{1}\dk{2} \; \frac{1}{(k_1^2-m_t^2) k_2^2 (k_1+k_2+p)^2} \,,
\end{align}
with $p^2=m_t^2$. The one- and two-loop strong coupling renormalisation constants are
\begin{align} \label{eq:Zas}
        \delta{Z}^{(1)}_{\as} &= -\frac{\beta_0}{\epsilon} \,, \\
	\dZ^{(2)}_{\as} & =  \frac{\beta_0^2}{\epsilon^2} - \frac{\beta_1}{2 \epsilon} \,, 
\end{align}
while those for the top-quark wave function, derived in the Feynman gauge, are~\cite{Melnikov:2000zc}
\begin{align} \label{eq:Ztop}
	\delta{Z}^{(1)}_{t} = {} & \frac{e^{\eps \gamma_E} \Gamma(1+\eps)}{(m_t^2)^\eps} \; C_F \left( -\frac{3}{\epsilon}
        - \frac{4}{1-2 \epsilon} \right) \,, \\
	\dZ^{(2)}_{t} = {} & \frac{e^{2\eps \gamma_E} \Gamma^2(1+\eps)}{(m_t^2)^{2\eps}} \; \bigg\lbrace  C_F^2 \, \bigg( \frac{9}{2 \epsilon^2} + \frac{51}{4 \epsilon}
    + \frac{433}{8} - 24 \zeta_3 + 96 \zeta_2 \log(2)-78 \zeta_2 \bigg) \nonumber \\
    & + C_F C_A \, \bigg( -\frac{11}{2 \epsilon^2} -\frac{101}{4 \epsilon}
	- \frac{803}{8} +12 \zeta_3 - 48 \zeta_2 \log(2) + 30 \zeta_2 \bigg) \\
    & + C_F T_F n_f \,  \bigg( \frac{2}{\eps^2} + \frac{9}{\eps} + 8 \zeta_2 + \frac{59}{2} \bigg) \nonumber
	\bigg\rbrace \,.
\end{align}
The beta-function coefficients are
\begin{align} \label{eq:BetaFunction}
	\beta_0 & = \frac{11}{3} C_A - \frac{4}{3} T_F n_f \,,\\ 
	\beta_1 &= \frac{34}{3} C_A^2 - \frac{20}{3} C_A T_F n_f + 4 C_F T_F n_f \,,
\end{align}
with
\begin{equation} \label{eq:Casimir}
	C_A = N_c \,, \qquad C_F = \frac{N_c}{2} \,, \qquad T_F = \frac{1}{2} \,.
\end{equation}
We note that a prefactor of $(4\pi)^{L \eps} e^{-L \eps \gamma_E}$ (where $L$ is the loop order) 
has been removed from the renormalisation constants $\dZ_m^{(L)}$, $\delta{Z}^{(L)}_{\as}$ and $\delta{Z}^{(L)}_{t}$
such that they match the prefactor convention of the $L$-loop amplitude specified 
in \cref{eq:colourdecompositionttggg,eq:colourdecompositionttqqg}. We also provide the anomalous dimensions required for the subtraction of IR singularities as specified 
in \cref{eq:GammaNg,eq:GammaNq,eq:gammaExpansion}~\cite{Becher:2009qa,Becher:2009kw}:
\begin{align} \label{eq:AnomDim}
        \gamma^{\mathrm{cusp}}_0 &= 4 \,, \\
	\gamma^{\mathrm{cusp}}_1 &= \left( \frac{268}{9} - \frac{4\pi^2}{3} \right) C_A - \frac{80}{9} T_F n_f \,, \\
        \gamma^{g}_0 &= - \beta_0 \,, \\
        \gamma^{g}_1 &= C_A^2 \left( - \frac{692}{27} + \frac{11}{18} \pi^2 + 2 \zeta_3  \right)
	                + C_A T_F n_f \left( \frac{256}{27} - \frac{2\pi^2}{9} \right) + 4 C_F T_F n_f \,, \\
        \gamma^{q}_0 &= -3 \, C_F \,, \\
	\gamma^{q}_1 &= C_F^2 \left( -\frac{3}{2} + 2\pi^2 - 24\zeta_3 \right)    
	                + C_F C_A \left( -\frac{961}{54}-\frac{11\pi^2}{6} + 26 \zeta_3 \right)  \nonumber \\
			& \quad + C_F T_F n_f \left( \frac{130}{27} + \frac{2\pi^2}{3} \right)\,, \\
        \gamma^{Q}_0 &= -2 \, C_F \,, \\
	\gamma^{Q}_1 &= C_F C_A \left( \frac{2\pi^2}{3}-\frac{98}{9}-4 \zeta_3 \right) + \frac{40}{9} C_F T_F n_f  \,.
\end{align}

\section{Scale dependence of the finite remainders}
\label{app:mudep}

We derived analytic expressions for the one- and two-loop leading colour finite remainders for $\ttggg$ and $\ttqqg$ with the renormalisation scale ($\mu_R$) set to $1$.
In this appendix we provide explicit formulae to restore their dependence on $\mu_R$. 
We define the $\mu_R$-restoring term $\delta F^{(L),i}_x$ for the term $F^{(L),i}_x$ of the partial finite remainder $\mathcal{F}^{(L)}_x$ defined in \cref{eq:FNcNfdecomposition} as
\begin{equation}
        F^{(L),i}_x(\mu_R^2) = F^{(L),i}_x(\mu_R^2=1) + \delta F^{(L),i}_x(\mu_R^2) \,.
\end{equation}
For $\ttggg$, the $\mu_R$-restoring terms are given by
\begin{align}
	\delta F^{(1),N_c}_{g}(\mu_R^2) & = F^{(0)}_g \, \bigg\lbrace \log(\mu_R^2) \, \big( -1 + g_1 \big)  - \frac{3}{2}\, \log^2(\mu_R^2) \bigg\rbrace \,, \\
	\delta F^{(1),n_f}_{g}(\mu_R^2) & = 0 \,, \\
	\delta F^{(2),N_c^2}_{g}(\mu_R^2) & = F^{(1),N_c}_g(\mu_R^2=1) \, \bigg\lbrace \log(\mu_R^2) \, \bigg( \frac{8}{3} + g_1 \bigg) 
	                                                                  - \frac{3}{2}\, \log^2(\mu_R^2) \bigg\rbrace  \nonumber \\
					  & \quad + F^{(0)}_g \, \bigg\lbrace  \log(\mu_R^2) \, \bigg[ -\frac{242}{9} + \frac{67 g_1}{9}  + \bigg(\frac{15}{2} - 2g_1 \bigg) \zeta_2 + \zeta_3 \bigg] \\
					& \hspace{1.7cm}   + \log^2(\mu_R^2) \, \bigg( -\frac{25}{2} + \frac{5 g_1}{6} + \frac{g_1^2}{2} + 3\zeta_2 \bigg) \nonumber \\
					& \hspace{1.7cm}   + \log^3(\mu_R^2) \, \bigg( -\frac{1}{3} - \frac{3 g_1}{2} \bigg) + \frac{9}{8}\log^4(\mu_R^2)  \bigg\rbrace \,, \nonumber \\
	\delta F^{(2),N_c n_f}_{g}(\mu_R^2) & = -\frac{2}{3}F^{(1),N_c}_g(\mu_R^2=1) \, \log(\mu_R^2) \nonumber \\ 
					  & \quad + F^{(1),n_f}_g(\mu_R^2=1) \, \bigg\lbrace \log(\mu_R^2) \, \bigg( \frac{8}{3} + g_1 \bigg) 
	                                                                  - \frac{3}{2}\, \log^2(\mu_R^2) \bigg\rbrace  \\
									  & \quad + F^{(0)}_g \, \bigg\lbrace  \log(\mu_R^2) \, \bigg( \frac{29}{9} - \frac{10 g_1}{9} - \zeta_2 \bigg)  \nonumber \\
					& \hspace{1.7cm} + \log^2(\mu_R^2) \, \bigg( 2 - \frac{g_1}{3} \bigg) 	+ \frac{1}{3}\log^3(\mu_R^2)  \bigg\rbrace \,, \nonumber \\
	 \delta F^{(2),n_f^2}_{g}(\mu_R^2) & = -\frac{2}{3} F^{(1),n_f}_g(\mu_R^2=1) \, \log(\mu_R^2)  \,,
\end{align}
while for $\ttqqg$ they read
\begin{align}
	\delta F^{(1),N_c}_{q}(\mu_R^2) & = F^{(0)}_q \, \bigg\lbrace \log(\mu_R^2) \, \bigg( \frac{7}{6} + g_2 \bigg)  -  \log^2(\mu_R^2) \bigg\rbrace \,, \\
	\delta F^{(1),n_f}_{q}(\mu_R^2) & = -\frac{2}{3}F^{(0)}_q \, \log(\mu_R^2) \,, \\
	\delta F^{(2),N_c^2}_{q}(\mu_R^2) & = F^{(1),N_c}_q(\mu_R^2=1) \, \bigg\lbrace \log(\mu_R^2) \, \bigg( \frac{29}{6} + g_2 \bigg) 
	                                                                  - \log^2(\mu_R^2) \bigg\rbrace  \nonumber \\
					  & \quad + F^{(0)}_q \, \bigg\lbrace  \log(\mu_R^2) \, \bigg[ -\frac{2275}{216} + \bigg(\frac{67}{9} - 2\zeta_2\bigg)g_2   + \frac{4}{3}\zeta_2 + 6\zeta_3 \bigg] \\
					& \hspace{1.7cm}   + \log^2(\mu_R^2) \, \bigg( -\frac{37}{8} + 3 g_2 + \frac{g_2^2}{2} + 2\zeta_2 \bigg) \nonumber \\
					& \hspace{1.7cm}   + \log^3(\mu_R^2) \, \bigg( -\frac{43}{18} - g_2 \bigg) + \frac{1}{2}\log^4(\mu_R^2)  \bigg\rbrace \,, \nonumber \\
	\delta F^{(2),N_c n_f}_{q}(\mu_R^2) & = -\frac{4}{3}F^{(1),N_c}_q(\mu_R^2=1) \, \log(\mu_R^2) \nonumber \\ 
					  & \quad + F^{(1),n_f}_q(\mu_R^2=1) \, \bigg\lbrace \log(\mu_R^2) \, \bigg( \frac{29}{6} + g_2 \bigg) 
	                                                                  - \log^2(\mu_R^2) \bigg\rbrace  \\
									  & \quad + F^{(0)}_q \, \bigg\lbrace  \log(\mu_R^2) \, \bigg( -\frac{71}{54} - \frac{10 g_2}{9} + \frac{2}{3}\zeta_2 \bigg)  \nonumber \\
					& \hspace{1.7cm} + \log^2(\mu_R^2) \, \bigg( -\frac{23}{18} - g_2 \bigg) + \frac{8}{9}\log^3(\mu_R^2)  \bigg\rbrace \,, \nonumber \\
	 \delta F^{(2),n_f^2}_{q}(\mu_R^2) & = -\frac{4}{3} F^{(1),n_f}_q(\mu_R^2=1) \, \log(\mu_R^2) + \frac{4}{9} F^{(0)}_q \, \log^2(\mu_R^2) \,.
\end{align}
Here, $g_1$ and $g_2$ are shorthands for the following linear combinations of weight-1 special functions,
\begin{align}
	g_1 & = -\pf^{(1)}_{1} + \pf^{(1)}_{3} + \pf^{(1)}_{4} + \pf^{(1)}_{5} + \pf^{(1)}_{6} \,, \\
	g_2 & = -\pf^{(1)}_{1} + \pf^{(1)}_{3} + \pf^{(1)}_{5} + \pf^{(1)}_{6} \,,
\end{align}
where we omit the argument $\vec{d}$.
These functions can be expressed in terms of logarithms of $\vec{d}$ as in eq.~(4.10) of ref.~\cite{Badger:2024dxo}.

\section{Contracted helicity finite remainders under external momentum permutations}
\label{app:camp_perm}

The computation of the colour dressed finite remainders entering the hard functions requires the evaluation of the partial 
finite remainders $\cF^{(L)}_x$ for different permutations of the external momenta 
according to~\cref{eq:colourdecompositionttgggfin,eq:colourdecompositionttqqgfin}. For a convenient bookkeeping when building the hard functions, 
we choose to use the same massive spinor structure in the decomposition of the partial amplitudes in~\cref{eq:massivespinordecomposition} (which holds similarly for the partial finite remainders) for all permutations, 
i.e.\ we keep the external momenta fixed in $\Psi_i$ and $\Omega_{ij}$.
Dropping the helicity ($h_3 h_4 h_5$) and $x$ indices for better readability, 
the partial finite remainder $\cF^{(L)}$ for an ordering of the external momenta $\momperm^\prime$ related to $\momperm = (12345)$ by a permutation $\sigma$
is given by
\begin{align}
	\cF^{(L)}(\momperm^\prime) &= \sum_{i=1}^{4} \Psi_i(\momperm) \, \cG^{(L)}_{i}(\momperm^\prime) \,, 
\end{align}
where
\begin{equation}
	\label{eq:formfactorspermuted}
	\cG^{(L)}_{i}(\momperm^\prime) = \sum_{j=1}^{4} \left(\Omega^{-1}(\momperm)\right)_{ij} \; \sum_{\mathrm{pol.}} \Psi^\dagger_i(\momperm) \,  \cF^{(L)}(\momperm^\prime)  \,.
\end{equation}
We evaluate $\cG^{(L)}_{i}(\momperm^\prime)$ from the contracted helicity finite remainders $\tilde{\cF}^{(L)}_{i}(\momperm^\prime)$,
that in turn we evaluate by permuting the external momenta in the analytic expression we computed for $\tilde{\cF}^{(L)}_{i}(\momperm)$.
To this end, we note that the contracted helicity finite remainder $\tilde{\cF}^{(L)}_{i}$ for the ordering $\momperm^\prime$ corresponds~to
\begin{equation} \label{eq:contractedampperm}
	\tilde{\cF}^{(L)}_{i}(\momperm^\prime) = \sum_{\mathrm{pol.}} \Psi^\dagger_i(\momperm^\prime) \,  \cF^{(L)}(\momperm^\prime) \,,
\end{equation}
with a mismatch in the argument of $\Psi_i^{\dagger}$ with respect to \cref{eq:formfactorspermuted}.
Therefore, in order to evaluate $\cG^{(L)}_{i}(\momperm^\prime) $, we need to rewrite $\sum_{\mathrm{pol.}} \Psi^\dagger_i(\momperm) \,  \cF^{(L)}(\momperm^\prime)$ in \cref{eq:formfactorspermuted} 
in terms of $\tilde{\cF}^{(L)}_{i}(\momperm^\prime)$.
Below we provide these relations for all the external momentum orderings $\momperm^\prime$ needed to build the hard functions:
\begin{itemize}

\item $\momperm_1^\prime = (12354)$ 
\begin{align}
\begin{aligned}
	\sum_{\mathrm{pol.}} \Psi^\dagger_1(\momperm) \,  \cF^{(L)}(\momperm^\prime_1) & = \tilde{\cF}^{(L)}_{1}(\momperm^\prime_1) \,, \\
	\sum_{\mathrm{pol.}} \Psi^\dagger_2(\momperm) \,  \cF^{(L)}(\momperm^\prime_1) & = \tilde{\cF}^{(L)}_{2}(\momperm^\prime_1) \,, \\
	\sum_{\mathrm{pol.}} \Psi^\dagger_3(\momperm) \,  \cF^{(L)}(\momperm^\prime_1) & = -\tilde{\cF}^{(L)}_{2}(\momperm^\prime_1)-\tilde{\cF}^{(L)}_{3}(\momperm^\prime_1) \,, \\
	\sum_{\mathrm{pol.}} \Psi^\dagger_4(\momperm) \,  \cF^{(L)}(\momperm^\prime_1) & = -\frac{2d_{12}}{m_t^2} \tilde{\cF}^{(L)}_{1}(\momperm^\prime_1)
	                                                                                  + 2 \tilde{\cF}^{(L)}_{2}(\momperm^\prime_1)
											  - \tilde{\cF}^{(L)}_{4}(\momperm^\prime_1) \,. 
\end{aligned}
\end{align}

\item $\momperm_2^\prime = (12435)$ 
\begin{align}
\begin{aligned}
	\sum_{\mathrm{pol.}} \Psi^\dagger_1(\momperm) \,  \cF^{(L)}(\momperm^\prime_2) & = \tilde{\cF}^{(L)}_{1}(\momperm^\prime_2) \,, \\
	\sum_{\mathrm{pol.}} \Psi^\dagger_2(\momperm) \,  \cF^{(L)}(\momperm^\prime_2) & = \tilde{\cF}^{(L)}_{3}(\momperm^\prime_2) \,, \\
	\sum_{\mathrm{pol.}} \Psi^\dagger_3(\momperm) \,  \cF^{(L)}(\momperm^\prime_2) & = \tilde{\cF}^{(L)}_{2}(\momperm^\prime_2) \,, \\
	\sum_{\mathrm{pol.}} \Psi^\dagger_4(\momperm) \,  \cF^{(L)}(\momperm^\prime_2) & = \frac{2d_{34}}{m_t^2} \tilde{\cF}^{(L)}_{1}(\momperm^\prime_2)
											  - \tilde{\cF}^{(L)}_{4}(\momperm^\prime_2) \,. 
\end{aligned}
\end{align}

\item $\momperm_3^\prime = (12453)$ 
\begin{align}
\begin{aligned}
	\sum_{\mathrm{pol.}} \Psi^\dagger_1(\momperm) \,  \cF^{(L)}(\momperm^\prime_3) & =   \tilde{\cF}^{(L)}_{1}(\momperm^\prime_3) \,, \\
	\sum_{\mathrm{pol.}} \Psi^\dagger_2(\momperm) \,  \cF^{(L)}(\momperm^\prime_3) & = - \tilde{\cF}^{(L)}_{2}(\momperm^\prime_3)
	                                                                                  - \tilde{\cF}^{(L)}_{3}(\momperm^\prime_3) \,, \\
	\sum_{\mathrm{pol.}} \Psi^\dagger_3(\momperm) \,  \cF^{(L)}(\momperm^\prime_3) & = \tilde{\cF}^{(L)}_{2}(\momperm^\prime_3) \,, \\
	\sum_{\mathrm{pol.}} \Psi^\dagger_4(\momperm) \,  \cF^{(L)}(\momperm^\prime_3) & = \frac{2(d_{15}-d_{23})}{m_t^2} \tilde{\cF}^{(L)}_{1}(\momperm^\prime_3)
											  - 2 \tilde{\cF}^{(L)}_{2}(\momperm^\prime_3) 
											  + \tilde{\cF}^{(L)}_{4}(\momperm^\prime_3) 
											  \,. 
\end{aligned}
\end{align}

\item $\momperm_4^\prime = (12534)$ 
\begin{align}
\begin{aligned}
	\sum_{\mathrm{pol.}} \Psi^\dagger_1(\momperm) \,  \cF^{(L)}(\momperm^\prime_4) & =   \tilde{\cF}^{(L)}_{1}(\momperm^\prime_4) \,, \\
	\sum_{\mathrm{pol.}} \Psi^\dagger_2(\momperm) \,  \cF^{(L)}(\momperm^\prime_4) & = \tilde{\cF}^{(L)}_{3}(\momperm^\prime_4) \,, \\
	\sum_{\mathrm{pol.}} \Psi^\dagger_3(\momperm) \,  \cF^{(L)}(\momperm^\prime_4) & = - \tilde{\cF}^{(L)}_{2}(\momperm^\prime_4)
	                                                                                  - \tilde{\cF}^{(L)}_{3}(\momperm^\prime_4) \,, \\
	\sum_{\mathrm{pol.}} \Psi^\dagger_4(\momperm) \,  \cF^{(L)}(\momperm^\prime_4) & = \frac{2(d_{34}+d_{45}-d_{12}-d_{23}-m_t^2)}{m_t^2} \, \tilde{\cF}^{(L)}_{1}(\momperm^\prime_4) \\
											& \phantom{=} \ + 2 \tilde{\cF}^{(L)}_{3}(\momperm^\prime_4) 
											  + \tilde{\cF}^{(L)}_{4}(\momperm^\prime_4) 
											  \,. 
\end{aligned}
\end{align}

\item $\momperm_5^\prime = (12543)$ 
\begin{align}
\begin{aligned}
	\sum_{\mathrm{pol.}} \Psi^\dagger_1(\momperm) \,  \cF^{(L)}(\momperm^\prime_5) & =   \tilde{\cF}^{(L)}_{1}(\momperm^\prime_5) \,, \\
	\sum_{\mathrm{pol.}} \Psi^\dagger_2(\momperm) \,  \cF^{(L)}(\momperm^\prime_5) & = - \tilde{\cF}^{(L)}_{2}(\momperm^\prime_5)
	                                                                                  - \tilde{\cF}^{(L)}_{3}(\momperm^\prime_5) \,, \\
	\sum_{\mathrm{pol.}} \Psi^\dagger_3(\momperm) \,  \cF^{(L)}(\momperm^\prime_5) & = \tilde{\cF}^{(L)}_{3}(\momperm^\prime_5) \,, \\
	\sum_{\mathrm{pol.}} \Psi^\dagger_4(\momperm) \,  \cF^{(L)}(\momperm^\prime_5) & = \frac{2(d_{15}-d_{23}+d_{45})}{m_t^2} \tilde{\cF}^{(L)}_{1}(\momperm^\prime_5)
											  - 2 \tilde{\cF}^{(L)}_{3}(\momperm^\prime_5) 
											  - \tilde{\cF}^{(L)}_{4}(\momperm^\prime_5) 
											  \,. 
\end{aligned}
\end{align}

\item $\momperm_6^\prime = (21345)$ 
\begin{align}
\begin{aligned}
	\sum_{\mathrm{pol.}} \Psi^\dagger_1(\momperm) \,  \cF^{(L)}(\momperm^\prime_6) & = \tilde{\cF}^{(L)}_{1}(\momperm^\prime_6) \,, \\
	\sum_{\mathrm{pol.}} \Psi^\dagger_2(\momperm) \,  \cF^{(L)}(\momperm^\prime_6) & = -\tilde{\cF}^{(L)}_{2}(\momperm^\prime_6) \,, \\
	\sum_{\mathrm{pol.}} \Psi^\dagger_3(\momperm) \,  \cF^{(L)}(\momperm^\prime_6) & = -\tilde{\cF}^{(L)}_{3}(\momperm^\prime_6) \,, \\
	\sum_{\mathrm{pol.}} \Psi^\dagger_4(\momperm) \,  \cF^{(L)}(\momperm^\prime_6) & = \frac{2d_{34}}{m_t^2} \tilde{\cF}^{(L)}_{1}(\momperm^\prime_6)
											  - \tilde{\cF}^{(L)}_{4}(\momperm^\prime_6) \,. 
\end{aligned}
\end{align}

\item $\momperm_7^\prime = (21354)$ 
\begin{align}
\begin{aligned}
	\sum_{\mathrm{pol.}} \Psi^\dagger_1(\momperm) \,  \cF^{(L)}(\momperm^\prime_7) & =  \tilde{\cF}^{(L)}_{1}(\momperm^\prime_7) \,, \\
	\sum_{\mathrm{pol.}} \Psi^\dagger_2(\momperm) \,  \cF^{(L)}(\momperm^\prime_7) & = -\tilde{\cF}^{(L)}_{2}(\momperm^\prime_7) \,, \\
	\sum_{\mathrm{pol.}} \Psi^\dagger_3(\momperm) \,  \cF^{(L)}(\momperm^\prime_7) & =   \tilde{\cF}^{(L)}_{2}(\momperm^\prime_7)
	                                                                                  + \tilde{\cF}^{(L)}_{3}(\momperm^\prime_7) \,, \\
	\sum_{\mathrm{pol.}} \Psi^\dagger_4(\momperm) \,  \cF^{(L)}(\momperm^\prime_7) & = \frac{2(-d_{12}-d_{23}+d_{34}+d_{45}-m_t^2)}{m_t^2} \tilde{\cF}^{(L)}_{1}(\momperm^\prime_7) \\
											& \quad - 2 \tilde{\cF}^{(L)}_{2}(\momperm^\prime_7) 
											  + \tilde{\cF}^{(L)}_{4}(\momperm^\prime_7) 
											  \,. 
\end{aligned}
\end{align}

\item $\momperm_8^\prime = (21435)$ 
\begin{align}
\begin{aligned}
	\sum_{\mathrm{pol.}} \Psi^\dagger_1(\momperm) \,  \cF^{(L)}(\momperm^\prime_8) & =   \tilde{\cF}^{(L)}_{1}(\momperm^\prime_8) \,, \\
	\sum_{\mathrm{pol.}} \Psi^\dagger_2(\momperm) \,  \cF^{(L)}(\momperm^\prime_8) & = - \tilde{\cF}^{(L)}_{3}(\momperm^\prime_8) \,, \\
	\sum_{\mathrm{pol.}} \Psi^\dagger_3(\momperm) \,  \cF^{(L)}(\momperm^\prime_8) & = - \tilde{\cF}^{(L)}_{2}(\momperm^\prime_8) \,, \\
	\sum_{\mathrm{pol.}} \Psi^\dagger_4(\momperm) \,  \cF^{(L)}(\momperm^\prime_8) & =   \tilde{\cF}^{(L)}_{4}(\momperm^\prime_8) \,.
\end{aligned}
\end{align}

\item $\momperm_9^\prime = (21453)$ 
\begin{align}
\begin{aligned}
	\sum_{\mathrm{pol.}} \Psi^\dagger_1(\momperm) \,  \cF^{(L)}(\momperm^\prime_9) & =   \tilde{\cF}^{(L)}_{1}(\momperm^\prime_9) \,, \\
	\sum_{\mathrm{pol.}} \Psi^\dagger_2(\momperm) \,  \cF^{(L)}(\momperm^\prime_9) & =   \tilde{\cF}^{(L)}_{2}(\momperm^\prime_9)
	                                                                                  + \tilde{\cF}^{(L)}_{3}(\momperm^\prime_9) \,, \\
	\sum_{\mathrm{pol.}} \Psi^\dagger_3(\momperm) \,  \cF^{(L)}(\momperm^\prime_9) & = - \tilde{\cF}^{(L)}_{2}(\momperm^\prime_9) \,, \\
	\sum_{\mathrm{pol.}} \Psi^\dagger_4(\momperm) \,  \cF^{(L)}(\momperm^\prime_9) & = \frac{2(d_{15}-d_{23}+d_{45})}{m_t^2} \tilde{\cF}^{(L)}_{1}(\momperm^\prime_9)
											  + 2 \tilde{\cF}^{(L)}_{2}(\momperm^\prime_9) 
											  - \tilde{\cF}^{(L)}_{4}(\momperm^\prime_9) 
											  \,. 
\end{aligned}
\end{align}

\item $\momperm_{10}^\prime = (21534)$ 
\begin{align}
\begin{aligned}
	\sum_{\mathrm{pol.}} \Psi^\dagger_1(\momperm) \,  \cF^{(L)}(\momperm^\prime_{10}) & = \tilde{\cF}^{(L)}_{1}(\momperm^\prime_{10}) \,, \\
	\sum_{\mathrm{pol.}} \Psi^\dagger_2(\momperm) \,  \cF^{(L)}(\momperm^\prime_{10}) & = -\tilde{\cF}^{(L)}_{3}(\momperm^\prime_{10}) \,, \\
	\sum_{\mathrm{pol.}} \Psi^\dagger_3(\momperm) \,  \cF^{(L)}(\momperm^\prime_{10}) & = \tilde{\cF}^{(L)}_{2}(\momperm^\prime_{10})
	                                                                                     + \tilde{\cF}^{(L)}_{3}(\momperm^\prime_{10}) \,, \\
	\sum_{\mathrm{pol.}} \Psi^\dagger_4(\momperm) \,  \cF^{(L)}(\momperm^\prime_{10}) & = -\frac{2d_{23}}{m_t^2} \tilde{\cF}^{(L)}_{1}(\momperm^\prime_{10})
											  - 2 \tilde{\cF}^{(L)}_{3}(\momperm^\prime_{10})
											  - \tilde{\cF}^{(L)}_{4}(\momperm^\prime_{10}) \,. 
\end{aligned}
\end{align}

\item $\momperm_{11}^\prime = (21543)$ 
\begin{align}
\begin{aligned}
	\sum_{\mathrm{pol.}} \Psi^\dagger_1(\momperm) \,  \cF^{(L)}(\momperm^\prime_{11}) & =   \tilde{\cF}^{(L)}_{1}(\momperm^\prime_{11}) \,, \\
	\sum_{\mathrm{pol.}} \Psi^\dagger_2(\momperm) \,  \cF^{(L)}(\momperm^\prime_{11}) & =   \tilde{\cF}^{(L)}_{2}(\momperm^\prime_{11})
											     + \tilde{\cF}^{(L)}_{3}(\momperm^\prime_{11}) \,, \\
	\sum_{\mathrm{pol.}} \Psi^\dagger_3(\momperm) \,  \cF^{(L)}(\momperm^\prime_{11}) & = - \tilde{\cF}^{(L)}_{3}(\momperm^\prime_{11}) \,, \\
	\sum_{\mathrm{pol.}} \Psi^\dagger_4(\momperm) \,  \cF^{(L)}(\momperm^\prime_{11}) & = \frac{2(d_{15}-d_{23})}{m_t^2} \tilde{\cF}^{(L)}_{1}(\momperm^\prime_{11})
											     + 2 \tilde{\cF}^{(L)}_{3}(\momperm^\prime_{11}) 
											     + \tilde{\cF}^{(L)}_{4}(\momperm^\prime_{11}) 
											     \,. 
\end{aligned}
\end{align}

\end{itemize}

\bibliographystyle{JHEP}
\bibliography{ttj_2L_lc}

\end{document}